\documentclass[preprint]{elsarticle}
\usepackage{subfigure}
\usepackage{tabularx}
\usepackage{url}
\usepackage{graphics, graphicx}

\newcounter{bla}

\journal{Computer Physics Communications}

\begin{document}
\begin{frontmatter}

\title{The MOLDY short-range molecular dynamics package}

\author[a]{G. J. Ackland\corref{graeme}} 
\author[a]{K. D'Mellow}
\author[a]{S. L. Daraszewicz}
\author[a]{D. J. Hepburn}
\author[a]{M.~Uhrin}
\author[a]{K.~Stratford}

\address[a]{School of Physics and Astronomy, The University of Edinburgh, King's Buildings,
  Edinburgh, EH9 3JZ}
\cortext[graeme]{Corresponding author.\\\textit{E-mail address:} g.j.ackland@ed.ac.uk}

\begin{abstract}

We describe a parallelised version of the MOLDY molecular dynamics
program. This Fortran code is aimed at systems which may be described by
short-range potentials and specifically those which may be addressed
with the embedded atom method. This includes a wide range of transition
metals and alloys. MOLDY provides a range of options in terms of the
molecular dynamics ensemble used and the boundary conditions which may
be applied. A number of standard potentials are provided, and the modular
structure of the code allows new potentials to be added easily. The code
is parallelised using OpenMP and can therefore be run on shared memory
systems, including modern multicore processors. Particular attention is
paid to the updates required in the main force loop, where
synchronisation is often required in OpenMP implementations of molecular
dynamics. We examine the performance of the parallel code in detail
and give some examples of applications to realistic problems, including
the dynamic compression of copper and carbon migration in an iron-carbon
alloy.

\begin{flushleft}
\end{flushleft}


\end{abstract}

\end{frontmatter}

\newpage
{\bf PROGRAM SUMMARY}

\begin{small}
\noindent
{\em Manuscript Title:} The MOLDY short-range molecular dynamics
package \\ {\em Authors:} G~J~Ackland K~D'Mellow S~L~Daraszewicz
D~J~Hepburn M~Uhrin K~Stratford \\ 
{\em Program Title:} MOLDY 
\\ {\em
Journal Reference:} \\ 
 {\em Catalogue identifier:} \\ 
 {\em Licensing provisions:} GNU General Public License Version 2 \\ 
 {\em Programming language:} Fortran 95 /  OpenMP 
\\ {\em Computer:} any \\ 
{\em Operating system:} any \\ 
{\em RAM:} 100 MB or more \\ 
{\em Number of processors used:} any \\ 
{\em Supplementary material:} Requires OpenMP for parallel execution
\\ 
{\em Classification:} 7.7 Other
Condensed Matter inc. Simulation of Liquids and Solids \\ {\em Nature
of problem:} Moldy addresses the problem of many atoms (of order
$10^6$) interacting via a classical interatomic potential on a
timescale of microseconds.  It is designed for problems where
statistics must be gathered over a number of equivalent runs, such as
measuring thermodynamic properities, diffusion, radiation damage,
fracture, twinning deformation, nucleation and growth of phase
transitions, sputtering etc. In the vast majority of materials, the
interactions are non-pairwise, and the code must be able to deal with
many-body forces.  \\ {\em Solution method:} Molecular dynamics
involves integrating Newton's equations of motion.  MOLDY uses verlet
(for good energy conservation) or predictor-corrector (for accurate
trajectories) algorithms.  It is parallelised using open MP.  It also
includes a static minimisation routine to find the lowest energy
structure.  Boundary conditions for surfaces, clusters, grain
boundaries, thermostat (Nose), barostat (Parrinello-Rahman), and
externally applied strain are provided.  The initial configuration can
be either a repeated unit cell or have all atoms given
explictly. Initial velocities are generated internally, but it is also
possible to specify the velocity of a particular atom.  A wide range
of interatomic force models are implemented, including embedded atom,
morse or Lennard Jones.  Thus the program is especially well suited to
calculations of metals.  \\ {\em Restrictions:} The code is designed
for short-ranged potentials, and there is no Ewald sum.  Thus for long
range interactions where all particles interact with all others, the
order-N scaling will fail.  Different interatomic potential forms
require recompilation of the code.  \\ {\em Additional comments: }
There is a set of associated open-source analysis software for
postprocessing and visualisation.  This includes local crystal
structure recognition and identification of topological defects.  \\
{\em Running time: } A set of test modules for running time are
provided.  The code scales as order N. The parallelisation shows
near-linear scaling with number of processors in a shared memory
environment.  A typical run of a few tens of nanometers for a few
nanoseconds will run on a timescale of days on a multiprocessor
desktop.


\newpage


\hspace{1pc}

\section{Introduction}

Classical molecular dynamics (MD), which solves Newton's laws of motion
for a system made up of atoms and/or molecules, is a powerful and
widely used tool to study both simple and complex systems. With the
power of modern computers, these systems can range in size from
a few thousand to many millions of atoms. Of particular interest
for modern hardware are codes which can take advantage of multicore
processors and increasingly accelerators, such as graphics
processing units (GPUs).

A number of well-established classical MD packages exist
which may be used to simulate very large systems in various contexts
[e.g., LAMMPS, DL\_POLY, NAMD]. These codes are able to address large
systems by means of domain decomposition and message passing, usually
via the Message Passing Interface (MPI). Some of these codes are important
for general purposes, while others are more fashioned for particular fields,
such as large biomolecules or proteins.

The code described here, MOLDY,
is one specifically developed with metals
and alloys in mind. This allows the code to be used for a wide range of
problems of practical interest, such as the properties of materials
undergoing deformation, defect propagation and radiation damage. It is not our intention to put forward a replacement for
codes of the type mentioned above, but to provide a complementary
option which allows users to simulate modestly large systems on
desktop machines. The increasing proliferation of multicore machines
makes OpenMP an ideal route to allow users less familiar with MPI the
chance to adapt the code incrementally without the additional
complexity involved with message passing. Further, the inclusion of
standard directives in future versions of OpenMP to distribute work
to accelerated hardware such as graphical processing units (GPUs)
would place the code in a good position to access these resources.

The code is a direct descendant of a Fortran code ``Moldy''
which was first developed at the United Kingdom Atomic Energy Research
Establishment \cite{gja,2}\footnote{Not to be confused with another
MD code of the same name: Refson \cite{1}.},     
with conventional constant volume and constant pressure ensembles and
deforming systems under constant stress using the extended Lagrangian
formulation
of Parrinello and Rahman \cite{3}. In Moldy, the atoms interact 
via short-range pairwise potentials and are free of bond constraints or
long-range interactions. These two simplifications are retained in the
current version. The code was extended to include the embedded atom
method \cite{4}, or Finnis-Sinclair potentials \cite{5}. 
This feature specifically aims the code
at the simulation of metals and alloys.

In retaining Fortran as the language of the updated version, we have
made the decision to restrict development to level of the Fortran
1995 standard. This is for reasons of
portability: compiler support for some features of later Fortran
standards is currently not complete. This means some design compromises
from a software engineering point of view. For example, an abstract
potential type, or class, might have been preferable for extensibility
and flexibility. Instead, at the moment, the potential is introduced
at compile time as a fixed module, so the code is recompiled for each different material. However, the
existing design does not preclude refactoring to make use of these
attractive language features in future. We hope the level of
compiler support for Fortran 2003 features will improve rapidly in
the near future.

In the following section we provide a brief background on molecular
dynamics which makes clear the issues addressed in the paper. Section
3 goes on to discuss the structure of the code and in particular
how the potentials should be represented. A number of different
potentials are included as examples. Importantly, this section
also discusses the
strategy for the parallelisation of the force loop using OpenMP.
Section~4 gives an overview of how the code is run in practice
and how the user deals with input and output.
Section~5 shows results for the parallel performance of the code
for a number of idealised problems and considers a
number of simple example applications. These applications present
an opportunity for visualisation and local crystal structure
determination, which are provided as part of the package. A summary
and conclusions are presented in Section~6.

\section{Theoretical background}

On account of the screening properties of the conduction electrons,
metals can be described by short range interatomic potential.  This
offer significant advantages for molecular dynamics simulations, in
particular since each atom interacts only with a finite number of
neighbours the computer time to calculate forces scales as $O(N)$
rather than $O(N^2)$ for generalised long ranged forces, or
$O(N^{1.5})$/$O(NlogN$ for Coulombic Ewald sums or fast multipole techniques.

For simulations of large numbers of atoms, it is essential that the
electronic degrees of freedom are integrated out, and forces depend
only on the atomic positions.  However, since the electrons are
delocalised, the forces cannot be simply pairwise interactions between
atoms. The delocalisation also suggests that an expansion in two-,
three, four-body terms may not converge quickly, if at all \cite{37}.  Since 1984 potentials
involving a function of some measure of the local density have become
the methods of choice.  The two most prominent are the Embedded Atom (EAM)
and the second-moment tight-binding methods, also called
Finnis-Sinclair (FS) \cite{5}. In these the cohesive energy is written as: 

\begin{equation}
U_{coh} = \sum_i\left( \sum_j V_{ij}(r_{ij}) + F_i[ \sum_j \phi_{ij}(r_{ij})
] \right ) \label{Ucoh} \end{equation}

where $V$ and $\phi$ are short-ranged pairwise potentials, and $F$ is a function of the sum of $\phi$'s.  The form of the second term means that an expansion of the potential in $n$-atom terms does not converge \cite{37}.

The interpretations of $\phi$'s are subtly different.  In EAM the $\phi_{ij} = \phi_j \ne \phi_i$ represents the
electron density from the atom $j$ at site $i$ - it depends only on
the type of atom at $j$.  In FS $\phi_{ij} = \phi_{ji}$ is the square
of the tight binding hopping integral between $j$ and $i$ - it depends
on both types of atom \cite{4,6}. In both cases $\phi$ is a function only of the
interatomic distance.
Similarly the EAM ``embedding function'' $F_i$ depends on the
atom at site $i$, while in the FS picture, $F$ is a universal function for a
canonical $d$-band, independent of atom types - the square root.
In practice, these distinctions only become important when more than
one atomic type is considered.

A number of potentials which appear to be different actually take the
same form as Eq.~\ref{Ucoh}.  These include the ``magnetic'' type
potentials\cite{DD,JNM}, rescaled potentials\cite{scale} and the Metallic-Covalent FeC potential
of Hepburn-Ackland \cite{7}. 

Similarly, simple pairwise potentials such as 
Morse or Lennard-Jones can be cast into this form by setting
$F(\rho)=0$.

The form of the potential allows one to define the ``energy of an
atom''.  This concept is both useful and meaningless.  It is useful
because deviations from the mean value can be used to highlight
defects or hotspots in a simulation, but meaningless because the physically meaningful quantities, total energy and forces, are independent of how the energy is partitioned
between atoms.

\subsection{Ensembles}

The code implements several different ensembles.

The ``Total Energy'' reported by MOLDY comprises only the kinetic
energy of the movable particles and the cohesive energy of the
system.  However the additional degrees of freedom of the extended
Lagrangian introduced by the thermostat and barostat can exchange
energy with the system.  Thus the only ensemble in which total energy
is conserved is NVEp, i.e. without barostat or thermostat.

\subsubsection{Molecular dynamics (microcanonical) NVEp}

The standard ``molecular dynamics'' ensemble is NVEp - constant
particle number, volume, energy and momentum (p=0).  The momentum
constraint distinguishes it slightly from microcanonical NVE. 

\subsubsection{Thermostat NVTp}

The Nose-Hoover \cite{8,9} thermostat can be applied to produce an  NVTp canonical ensemble. 
This couples the system to an external heat bath.  The temperature is defined
by the kinetic energy via the relation:

\[ \sum_i \frac{\bf p_i^2}{2m_i} =   \frac{3N-3}{2}k_B T \]

where $k_B$ is the Boltzmann constant, $m_i$ and {$\bf p_i$} are the
particle masses and momenta and (3N-3) is the number of degrees of
freedom once total momentum is conserved.

The Nose-Hoover thermostat is an extended Hamiltonian, which can be interpreted as an external
heat reservoir, with an additional degree of freedom. This heat
reservoir controls the temperature through
exchange of kinetic energy with the system.  To define a heat bath kinetic
energy we introduce an effective mass $Q_s$ and momentum
$\pi_s$.
 The equations of motion are then

\begin{displaymath} \frac{d \bf r_i}{d t}=\frac{ \bf p_i}{m_i}, \end{displaymath}

\begin{displaymath} \frac{d \bf p_i}{dt}=-\frac{\partial U}{\partial \bf q _i}- \pi_s {\bf p_i}, \end{displaymath}

\begin{displaymath}  Q_s\frac{d \pi_s}{d t} =\sum_i \frac{{\bf p_i}^2}{2m_i} - (3N-3)kT .
 \end{displaymath}

Thus $\pi$ is the degree of freedom associated with the thermostat,
effectively a thermodynamic friction and $ Q_s$ is ``thermal
inertia'', which determines the rate of heat transfer and how closely
the temperature is maintained near to the target temperature
$T_{opt}$.  The value of $Q_s$ depends on the
heat conduction in the system and must be specified in the input to
the code.  It is possible to choose $Q_s$ to reproduce missing effects
such as electronic heat conduction.

\subsubsection{Barostat NPTp}

The Parrinello-Rahman method \cite{3} is implemented for simulations in the
NPEp (or with Nose, NPTp) ensemble.  There are a number of anomalies
in that paper\cite{NoseKlein,Cleveland,Wentzkovitch}, including lack of rotational invariance and missing
cross terms in the derivatives, and these are reproduced faithfully by
the code.

The positions {\bf $r_i$} are written as a product of fractional
coordinates ${\bf x_i}$ and a 3$\times$3 ``boxmatrix'' $h = [\bf{a,b,c}]$ defining the simulation volume

\[  {\bf r_i} =  h {\bf x_i} \]

from which a strain matrix with respect to a reference structure $h_0$
is defined by

\[ \epsilon = \frac{1}{2}({h'_0}^{-1}h'hh_0^{-1}-1) \]

where prime denotes the transpose.
The boxmatrix introduces nine additional degrees of freedom, three
stretches, three shears and three rotations.  Equations of motion for
these degrees of freedom come from the stresses on the supercell.

The strain definition is non-unique, but is  sensible in the limit of small
strain.  The Parrinello-Rahman Lagrangian contains the kinetic and
potential energy of the particles, pressure times volume, a fictitious
``Kinetic Energy'' associated with the boxmatrix and a complex term
arising from the non-hydrostatic stresses.

\[ L = \frac{1}{2}\sum_i m_i\dot{\bf x'}_ih'h\dot{\bf x}_i
- U_{coh}- P(\Omega-\Omega_0)
+ \frac{1}{2}W {\rm Tr}\dot{h}'\dot{h}
-   \frac{1}{2} {\rm Tr}[h_0^{-1}({\bf S}-P){h'}_0\Omega_0h'h.  \]

Where $\Omega, \Omega_0$ are the volume and unstrained volume, respectively; P is
the external hydrostatic pressure, {\bf S} is the external stress.  A constant
term $\Omega$Tr$({\bf S}-P)$ is ignored.

The scalar $W$ is the equivalent of a mass associated with the box degrees of
freedom. It determines how rapidly the box changes shape in response
to stress and can be related loosely to elastic constants. Typically
it takes a value of similar order of magnitude to the sum of the
atomic masses. From all this analysis, the equation of motion for the
boxmatrix degrees of freedom is

\[ W\ddot{h} = ({\bf \pi}-P)\sigma -h[h_0^{-1}({\bf S}-P){h'}_0\Omega_0] \]

where ${\bf \pi}$ is the internal stress tensor from the kinetic energy
and virial and ${\sigma}$ is the reciprocal space equivalent of the boxmatrix

\[ \sigma = \{ b \times c, c \times a, a \times c \}.\]

\subsection{Boundary conditions}

The code offers a range of different boundary conditions. For infinite
simulations, the simulation cell is repeated periodically, and the
minimum image convention for forces is applied where 
for purposes of eq.\ref{Ucoh} $r_{ij}$ refers to the shortest vector between atom $i$ and either $j$ or any of its periodic images.

Each
atomic trajectory is followed faithfully, there is no explicit
wraparound of the atoms (i.e position is stored as {\bf r }=$(x,y,z)$
rather than {\bf x }=$(x (mod\, a) ,y (mod\, b) ,z (mod\, c))$.  This allows
diffusion calculation to be done reliably.

As an alternative, free boundary conditions are also possible, where
there are no repeated cell or periodic boundary conditions.

For surface calculations, free surface in the $z$ direction can be
combined with minimum image convention in the $x$ and $y$ directions.
For surface calculation in crystals there is a surface tension
(different from the surface energy) \cite{Thetford}.
Ideally, one would like to have
an infinitely thick slab so this 2D tension would be balanced by an
infinitesimal 3D strain: in practice an infinite slab is not possible,
and the correct boundary condition to simulate macroscopic material is to impose constant supercell lengths with
(NVT)
minimum image forces in $x-y$ and free boundaries in $z$.  Relaxation
perpendicular to and within the surface is then still possible.

For interface calculation (e.g. grain boundaries) the interface may
 expand, so the boundary condition should be constant (zero) stress
 perpendicular to the boundary and constant strain within the
 boundary.  Once again this will generate internal stresses in the
 simulation, which in reality are balanced by infinitesimal strains in the bulk.

Finite strain rate boundary conditions can be applied using the {\it
strainloops} variable.  This multiplies the boxmatrix by the
strain matrix every {\it nsteps}, and continues the calculation for a
total of {\it strainloops*nsteps} timesteps.  This enables an external
shear or compression to be applied stepwise to the system.  By
reducing the values of {\it nsteps} and {\it straintensor} while
increasing {\it strainloops} it is possible to make this shear as
smooth as one wishes.

A similar method can be used to achieve finite temperature gradient:
the {\it tempsp} parameter increments the target temperature every
{\it nsteps} timesteps.  This then allows the thermostat to add or
extract the additional energy.

\subsection{Time integration}

The equations of motion for atomic and box degrees of freedom can be
integrated using either of two methods: velocity Verlet\cite{verlet}
or third-order predictor-corrector\cite{gear}.  Velocity Verlet is
time-reversible, and so tends to give better energy conservation (see
Section~\ref{section:integration_error}). By contrast, the third-order
predictor-corrector gives more accurate atomic trajectories.

\subsection{Fixed atoms}
In some case it is desirable to fix atoms so that they do not move
under external forcing.  MOLDY fixes the fractional coordinates of
all atoms with $atomic\_mass=0$, at constant values.  By applying an
external strain to the boxmatrix, it is possible to move rigid
blocks of atoms around to apply external strains to the system.  There
are possible consequences of this: any unusual behaviour at the
interface between fixed and not fixed atoms is likely to be
unphysical, and phonons or sound waves are reflected from such
surfaces.  The stress calculation ignores any forces
acting between fixed atoms.

\subsection{Velocity initialisation}

At the start of the calculation all atoms are assigned velocities
drawn from a uniform random distribution and scaled such that the
kinetic energy is equal to the target temperature (as input) and the
total momentum is zero.  If all atoms are started on their
crystallographic positions, this means that the initial state has all
its energy in kinetic degrees of freedom.  After NVEp equilibration,
the temperature will typically drop to one half of the initial value,
while in NVTp the thermostat will supply energy to maintain the
required temperature.  To ensure that there is some initial variation
in potential energy, it is possible to displace all atoms randomly
using the {\it dsp} parameter.

A special parameter group {\it pka,pkavx,pkavy,pkavz }  and {\it Epka }, allows the user to specify a single particle's initial kinetic energy and direction of motion.  This enables radiation damage simulations to be
done by defining the incoming or primary knock-on atom.

\subsection{Thermodynamic Quantities}

MOLDY can calculate a number of thermodynamic quantities.  The
simplest of these are quantities defined for a microstate, number of
degrees of freedom, coordinates and their time derivatives:
$\tilde{N},{\bf x_i,
\dot{x_i}},h$; Temperature
\[ T= \frac{1}{\tilde{N}k_B}\sum \frac{1}{2}\sum_i m_i\dot{\bf x'}_ih'h\dot{\bf x}_i;  \]
Potential Energy, $U_{coh}$; kinetic energy $KE=3\tilde{N}k_BT/2$; total energy $TE=U_{coh}+KE$, volume $V=det(h)$ and
enthalpy $H=TE+PV$, where P is the external pressure.

More complicated are those quantities which can be defined through
the various fluctuation theorems\cite{AllenTildesley}.  For example the specific heat capacities

\[ c_V = \frac{\tilde{N}k_B}{2}+\frac{\langle \delta U_{coh}^2 \rangle_{NVT}}{k_bT^2} \]

\[  c_P=\frac{\tilde{N}k_B\langle T \rangle_{NPE}}
        {2\langle T \rangle_{NPE}-3\tilde{N}\langle \delta T^2 \rangle_{NPE}} \]

the compressibility
\[    \kappa=  \frac{\langle \delta V^2 \rangle_{NPE}}
          { \langle V \rangle_{NPE} k_B\langle T\rangle}_{NPE} \]

and the thermal expansion
\[\alpha=\left (
\frac{2c_P}
{\tilde{N}k_B\langle T \rangle_{NPE}^2}\right ) \frac{
   \langle \delta V\delta U_{coh}\rangle_{NPE} + P\langle \delta V^2 \rangle_{NPE}}{\langle V\rangle_{NPE}}
\]

where  $\delta A$ is the variation from the average values,  $\langle \delta A^2\rangle$ found by
keeping a running average of $\langle A^2 \rangle-\langle A \rangle^2$.

These quantities can be calculated as running averages over a
molecular dynamics simulation.  However great care has to be taken to
obtain good results. The simulation must be correctly sampling the
ensemble throughout, so proper equilibration is essential.  Secondly,
these quantities converge slowly with time, so extended runs are
needed.  In each case it is possible to calculate the same quantity
from a series of runs varying the external T, V or P, measuring
the mean values of microstate quantities, and obtaining $c_V$, $c_P$, $\alpha$ or $\kappa$ from the slope.

\subsection{Units}
\label{units}

MOLDY defines the following system of units, naturally applicable to
molecular dynamics, and uses these internally: length in $\AA$, mass in
atomic mass units ($1u=1.66053886 \times 10^{-27}kg$) and energy in electron
volts. The units of time are constructed as a combination of these base
units with dimensions $(M \times L^2/E)^{1/2}$. As this gives an unusual
number $(\approx 1.018$x$10^{-14}s)$ we convert all input from femtoseconds to
model units and output from model units to femtoseconds for human
consumption. Hence timescales are always considered by the user in fs.

Similarly, the internal units of pressure are constructed as
a combination of the base units with dimensions $(E/L^3$).
However, for convenience, all input and output concerning
pressure are in units of GPa. The necessary conversions to and
from internal units are done automatically.  Similarly, temperature is input and output in Kelvin.

\section{Potentials and the Force Calculation}

\subsection{The potential interface}

This module provides the interface to the potential. The material
choice is made at compile-time (using preprocessor directives in the $Makefile$). By
design, the core code only interacts with the public interface of
\verb|potential_m| and not the actual potential as supplied by the
relevant material module.  Therefore, each material module must
conform to the well defined interface.

Public routines fall into two categories in a material module:
those of the potentials themselves and those of inquiry routines. A material module must provide public functions for the pair potential
$V$, the cohesive potential $\phi$, the embedding function $E$, and their derivatives:

\begin{verbatim}
 function vee_src(r, na1, na2)
 function dvee_src(r, na1, na2)
 function phi_src(r, na1, na2)
 function dphi_src(r, na1, na2)
 function emb_src(rho, na1)
 function demb_src(rho, na1)
\end{verbatim}
where the arguments for the pairwise potential are of the following type:

\begin{verbatim}
    real (kind_wp) :: vee_src       !< result (eV)
    real (kind_wp), intent(in) :: r !< separation (angstrom)
    integer, intent(in) :: na1      !< atomic number atom 1
    integer, intent(in) :: na2      !< atomic number atom 2
\end{verbatim}

The $V$ and $\phi$ potentials accept two species arguments, allowing
one to specify asymmetric
potentials, {\it e.g.} $\phi_{ij}(r) \neq
\phi_{ji}(r)$, $V_{ij}(r) \neq V_{ji}(r)$. The
embedding function is only a function of $\rho$ for a given species.

In addition to the $V$, $\phi$, and $E$ potentials, any material
 module must provide two inquiry routines:

\begin{verbatim}
  subroutine get_supported_potential_range(rmin,rmax)
    real(kind_wp), intent(OUT) :: rmin, rmax !< range in angstroms
\end{verbatim}

This returns the minimum and maximum range over which the potential
acts. MOLDY uses this information when setting link-cell and neighbour
list sizes and in deciding whether to calculate the potential for a
particle pair. The implementation of this routine is not specified and
is free to the user. Examples are found in the supplied material modules.

\begin{verbatim}
  subroutine check_supported_atomic_numbers(species_number,spna,ierror)
    integer, intent(in) :: species_number       !< number of species
    integer, intent(in) :: spna(species_number) !< atomic numbers of the species
    integer, intent(out):: ierror               !< return error code
\end{verbatim}

At startup, this routine checks that the available material module can
support the particles contained within the simulation's system file,
and should return non-zero \verb|ierror| if this is not the
case. MOLDY will halt execution before the simulation begins, and
report the condition. In the case of the {\it generic} potential module, this routine
loads the required potential coefficients from coefficients files or
returns an error if the files are not found.


\label{Available potentials}
\subsubsection{Available potentials}

The process of choosing a material module is one which might employ function pointers, available in Fortran 2003 onwards. However,
as we adhere to Fortran 95 standards for portability, we have ruled this out
as an option. Consequently, a material module is chosen at compile
time, suitable for the simulation to be performed. There are two
stages required to introduce an additional material module to the code:

\begin{enumerate}
\item Create a material module conformant to the potential interface.
\item Edit a single line in \verb|potential_m.F90| to optionally ``use'' this module, e.g.:

\begin{verbatim}
  !Material Module choice - use one material module only.
  use generic_atvf
\end{verbatim}
\end{enumerate}

For convenience, one can make use of the \verb|MATERIAL|  variable
to allow the choice to be made from within the {\it Makefile}.

The above forms the core code of MOLDY. The code's design makes it easy
to change potentials without any changes to the main
code. Consequently, we have provided a potentials module, which delivers an abstract interface through which the code accesses a
selected set of potential functions from {\it material module}. A material module can be written or modified by the user
and is required at compile-time, however it must conform to
the defined potential interface. The following material modules are provided in the initial release:\\
\\
\verb|iron_carbon| --- complete asymmetric potential set for Fe-C simulations \cite{7,31}\\
\verb|zirconium| --- single species Zr hcp potential \cite{30}\\
\verb|pairpot| --- multispecies Lennard-Jones potential with parameters read at runtime \\
\verb|morse| --- multispecies Morse potential with parameters read at runtime \\
\verb|generic_atvf| --- generic potential loading module to load
cubic spline potential coefficients from disk at runtime \cite{29}\\

The generic potentials currently offer single species potentials for
Al, Cs, K, Li, Mo, Na, Nb, Rb, Ta, V, W, Cu, Au, Ag, Ni, Ti, Zr, and
Pt. In addition, Ni-Al cross species potentials are included. Each
potential is specified in a coefficients file. An example of a
coefficients file is given here:

\begin{verbatim}
# ATVF format Ni/Al fitted to apb001 and elastic constants #
13             # Principal Atomic number
28             # Second Atomic number (used only for cross potential)
4              # Number of coefficients (vee)
1              # Number of coefficients (phi)
0.0            # Minimum potential radius
4.35174        # Maximum potential radius
-0.6469208776  1.1392692848  -0.6655106072  1.4680219296 # a_k coefficients (v)
4.35174 4.24473 3.88803 2.96061                          # r_k coefficients (v)
0.0            # a_k coefficients (phi)
1.0            # r_k coefficients (phi)
# END OF INPUT #
\end{verbatim}

Coefficients files must conform to the naming convention which MOLDY
employs: {\it pot\_XXX\_YYY.in} where {\it XXX} and {\it YYY} are the
atomic numbers of the two species involved suitably prefixed with
zeros. For example, the Al potential (single species) is {\it
  pot\_013\_013.in}, whereas the Ni-Al cross potential is provided by
both {\it pot\_013\_028.in} and {\it pot\_028\_013.in}. Additional
materials can be provided by supplying the appropriate potential files.

For Morse and Lennard Jones potentials it is still necessary to
specify an atomic number, which will determine the mass.  Although these potentials cannot represent specific materials, units are still required.  
It is recommended that parameters be chosen such that lattice parameters and binding energies lie in the $\AA$ and $eV$ range as MOLDY makes some internal checks for unphysical values.

\subsection{The Force calculation}

\subsubsection{The Neighbourlist calculation}

MOLDY uses a hybrid neighbour-list/link-cell method\cite{FH,gja}. The
link-cell method, or LCM was designed for a large number of
particles, because when using LCM the calculation scales as $O(N)$. When a
neighbour list is used in the traditional way, a calculation tends to
scale as $O(N)$ at each timestep at which the list is consulted, but
as $O(N^2)$ at each timestep in which the list is updated. The
hybrid method, removes the remaining
$O(N^2)$ scaling because the neighbour
list is rebuilt but from neighbours within adjacent
link cells, and the link cells are fixed in fractional coordinate space.

In a dynamic simulation the link cells in the NL-LCM must be large
enough to include all neighbours which might come into range of an
atom within the cell/list update period. MOLDY ensures that link cells
are a minimum size according to the maximum range of the acting
potentials, plus a user-defined padding size \verb|rpad| given in the
parameter file.

The neighbour list is calculated relatively infrequently, but for
large systems, this can be costly. The neighbour list is the second most
intensive task after the force calculation, and in tests we find the
neighbourlist quickly becomes a bottleneck. The calculation is
implemented as a nested loop over link-cells and their adjacent cells.
As calculation of the neighbour list does not change the state of the
system, the lists for each particle can be calculated independently. We find that applying a simple loop over a single link-cell index works effectively.

\subsubsection{OpenMP Parallelisation}

The code has been parallelised for shared-memory systems using OpenMP.\footnote{
in fact, we used OpenMP Version 2.4}  This type of system is often found in
modern hardware, as multi-core processors have become increasingly
standard, and look to increase further in both core number and
ubiquity. The two targets for parallelisation of MOLDY are the force
and neighbourlist calculations.

For reasons of efficiency, MOLDY maintains an asymmetric neighbour
list, such that for two atoms $i$ and $j$, if $j$ is in the neighbour
list of $i$, the reverse is not true. This allows the force $F_{ij}$
to be calculated once and applied to both particles $i$ and $j$,
without double counting. We adopt a particle-based parallelisation, as
this promises the most even load-balance and maps sensibly onto the
particle based loops. This implies that a particular OpenMP
thread will have assigned to it a set of particles, for which it is
responsible. In what follows, this thread is referred to as the parent
of these particles. To calculate the total force on a particle $i$, we
sum the force components $F_{ij}$ for all the neighbours of
$i$. The  difficulty in this approach
arises when the force  $F_{ij}$ is also to be applied to $i$'s
neighbouring particle $j$, where $j$ does not belong to the same
OpenMP thread as $i$. There is a real and unacceptable risk that two
threads will attempt to update the force on $j$ at the same time -
those threads being the parent thread of $i$ whose neighbour is $j$,
and the parent thread of $j$ itself. In order to avoid this behaviour,
synchronisation must be employed, such that only one thread may update
the cumulative force on $j$ at a time. However, a simple
synchronisation like atomic updates or a critical region, proves too
costly for the force calculation, as the necessary synchronisation
introduces a substantial overhead to this intensive region of code. It
should be noted that it is possible to avoid any kind of
synchronisation by maintaining a symmetric neighbourlist, at the cost
of having to calculate $F_{ij}$ twice. This factor of two penalty is
only worth taking if a synchronisation mechanism results in less than
50\% relative efficiency.

To combat this, we have established a mechanism that results in
far fewer synchronisation calls. We adopt the approach of locking
access to particles within defined regions of the simulation volume,
encompassing the range of influence of a particular particle - by
design, the link-cell scale. Establishing locks on the link-cells
allows us to lock/unlock a larger region of space, a correspondingly
infrequent number of times. This manifests as a larger wait time should a
non-parental thread need to access a particle. By ordering the
particles array by increasing link-cell index, it is possible to space
out the regions of the simulation volume being acted on by separate
threads. This generally minimises the number of times different
threads will be working within range of each other, and hence,
minimises the time spent waiting for locks. This mechanism is designed
to perform well for larger problem sizes.

From this basis, three schemes were considered, as follows:\\
\\
A:  The parent thread locks the link-cell containing its child particle $i$, and also the link-cell of the neighbour particle $j$. Link-cells can be locked and unlocked as infrequently as the link-cells change.
\\
B:  The parent thread locks each link-cell within the sphere of influence of a
particle, and only releases these locks when the link-cell or the
current particle $i$ changes from that of the previously considered
particle $i-1$. This involves locking 8 cells per thread.
\\
C: The parent thread locks the link cell of the neighbouring particle
$j$ only, and releases when the neighbouring cell changes from that of the
previously considered neighbour $j-1$. This involves locking only
one cell per thread, and calculating a private accumulator of the
force on the child particle $i$, updating this at the end with a
similar lock/unlock on $i$'s cell.\\

Scheme A is the simplest, but if not careful, schemes like this can
result in deadlock if two threads are visiting neighbours in the
link-cells of each other's child particles. Scheme B is an overly
conservative approach, locking more of the simulation volume than is
strictly necessary at one time. This scheme also requires extra care
to be taken over locking order, to avoid deadlocking
situations. Scheme C has been adopted, as it benefits from
locking only one link-cell at a time, which eliminates the
possibility of deadlock, and decreases the number and duration of waits.

For MOLDY, the neighbour list is constructed from the link cells so that
atoms in the same link cell are contiguous.  This means that in the loop over neighbours, at most 8 lock/unlock commands need to be sent.  
Tests showed that this is considerably faster than double-calculation
of the potential for large systems, albeit at the cost of some coding
complexity.  The scheme is illustrated in the pseudocode snippet below.
\begin{verbatim}
!$OMP PARALLEL

neighbour_link_cell = 0 !! (null value = neighbour link-cell is not set)

!$OMP DO
  !! calculate force on particles and their neighbours
  particleloop: do i = 1, number of particles

    ! Temporary force accumulator for particle i
    force_i = 0

    neighbourloop: do j = 1, number of neighbours of particle i
       
      ! Global index of i particle's j neighbour
      neighbour_idx = neighbours(j,i)
          
      force_ij = calculate force between i and j
            
      !! set/reset region locks when changing neighbour link cell
      if (linked_cell_of(neighbour_idx) .ne. neighbour_link_cell) then
        !! If we hold a lock on an old link cell, unset it
        if (neighbour_link_cell .gt. 0) then 
          call omp_unset_lock(lock(neighbour_link_cell))
        end if
        !! set neighbour link cell to the current link-cell and lock it
        neighbour_link_cell = linked_cell_of(neighbour_idx)  
        call omp_set_lock(lock(neighbour_link_cell))
      end if
          
      !! update force of particle's neighbour
      force(neighbour_idx) += force_ij
          
      !! update temporary force accumulator for particle i
      !! do not update force on i yet.         
      force_i += force_ij
          
      end do neighbourloop

    !! set/reset region locks when updating own particle
    if (linked_cell_of(i) .ne. neighbour_link_cell) then
      call omp_unset_lock(lock(neighbour_link_cell))
      !!set neighlc to the current link-cell
      neighbour_link_cell = linked_cell_of(i)
      call omp_set_lock(lock(neighbour_link_cell))
    end if
    force(i) -= force_ij

  end do particleloop
 
!$OMP END DO NOWAIT
!! release all locks
call omp_unset_lock(lock(neighbour_link_cell))
\end{verbatim}

\section{Running the Code}

\subsection{Input and Output files}

MOLDY utilises two main input files: the {\it parameter} file, and the
{\it system} file. These two are sufficient to describe a
physical system and the  features of a simulation.

The main output files MOLDY creates are the default output
file (equivalent to stdout), and the unformatted checkpoint/restart
file. Input and output filenames can be specified by the user in the main parameter file.

\subsubsection{The main parameter file, {\it params.in}} \label{parameterfile}

The main input file is {\it params.in}, and must always have this
name. This file contains all the parameters of the simulation,
including specification of other file names, physical parameters of the simulation,
verification parameters for the system file, and non-physical simulation parameters
governing flow control and loop counters. The parameter file contains key-value pairs. Comments can be placed at any point
at the end of a line or on separate lines. Many parameters have
defaults, in which case specification is not necessary. The parameter list is given in Tab.~\ref{parametertable}.

\begin{table}
\caption{Defined key-value pair parameters and their defaults}\label{parametertable}
\begin{tabularx}{\textwidth}{ll X}
Keyword & Default & Description\\
\hline
file\_system & system.in &  File describing atomic positions in system (the system file) \\
file\_textout & moldin.out & File for text output \\
file\_checkpointread & - & Input checkpoint filename \\
file\_checkpointwrite & - &       Output checkpoint filename \\
file\_dumpx1  & -  &  File to output per iteration data\\
boxmass & - & Mass of the box in units of A.M.U. per constituent particle \\
deltat  & - & Timestep in fs \\
rpad    & 0.0 & Padding thickness over potential cutoff in \AA (for link cell) \\
temprq  & - & Required temperature in the MD in Kelvin \\
press   & 0.0 & Required external pressure in GPa \\
ivol    & -  & Switch defining volume behaviour of the system \\
iquen   & - & Switch between molecular dynamics and molecular statics \\
nose    & 0.0 & Softness of damping in the Nose thermostat (0.0: no damping) \\
dsp     & 0.0 & Random displacement from input particle positions (in \AA) \\
pka     & 0 & Prints out position of a single chosen atom each nprint steps \\
Epka    & 0 & Initial energy of atom Epka \\
pkavx,pkavy,pkavz     & 0,0,0 & Direction of motion of pka  \\
straintensor  & $\delta_{ij}$&  Strain tensor in one unrolled 9 element value\\
nm      & - & The number of particles in the simulation \\
nspec   & - & The number of distinct species in the simulation\\
nsteps    & 0 & Number of timesteps to be done in a run \\
nbrupdate & - & Frequency of updating list of neighbours (MD steps) \\
strainloops    & 1 & Number of loops of applying the strain tensor \\
restart   & 0 & Switch defining whether a restart file is read at startup\\
nprint    & 100 & Frequency of printing of thermodynamic averages (MD steps)\\
nchkpt    & -1 & Frequency with which to checkpoint (MD steps)\\
nposav    & 0 & Prints sys\_avs.out: positions averaged over final nposav timesteps\\
nnbrs     & 150 & Maximum size of the neighbour list \\
tjob      & - & Maximum execution time in seconds \\
tfinalise & - & Time reserved for finalising the job \\
iverlet   & 0 & Algorithm choice (0=predictor-corrector, 1=verlet) \\
prevsteps & 0 & Number of timesteps done previously ({\it e.g.,} if restarting) \\
lastprint & 0 & Number of timesteps since last printing of run averages \\
lastchkpt & 0 & Number of timesteps since last checkpoint \\
dumpx1  & .False. & Flag to write per-iteration data to file   \\
write\_{}rdf & .False. & Flag to write the final radial distribution function to file on completion \\
\end{tabularx}
\end{table}

To get started with the code, some parameters of note are the
following:

\begin{description}

\item[boxmass]{In Parrinello-Rahman dynamics, the box itself is
  dynamically evolved. The box mass determines the magnitude of the reaction
  of the box as the system evolves. The mass is specified in atomic
  mass units per particle.}
\item[rpad]{The neighbour list is constructed from particles within
  the potential range. A padding distance is added to maintain the
  completeness of the neighbourlist as particles move into and out
  of the potential range between neighbourlist updates. A rule of thumb for the
  padding distance is the expected maximum drift of particles over the
  course of the time between neighbour-list recalculation. This will be system dependent and vary a lot between, e.g. solid and liquid.}
\item[ivol]{This integer parameter is a switch which controls the MD details. Accepted values are (0) constant pressure; (1) constant
  volume; (2) free surface on z, constant volume on x,y; (3) cluster: free surface on x,y,z,
  (4) grain boundary: constant volume on x,y, constant pressure on z.
(5) Pillar compression: free-surface-on-xy, constant-volume-on-z.
           }
\item[iquen]{Integer parameter to choose between molecular dynamics,
  molecular statics (quenching) or ``boxquenching''.}
\item[nose]{The Nose-Hoover thermostat parameter; related to the $Q_s$
defined above, large values correspond to slow relaxation.}
\item[straintensor]{An external strain can be applied on the
simulation volume (box) after construction. This is a 3x3
matrix, but can be specified here unrolled as nine values, {\it
  i.e.} $S_{xx}\ S_{xy}\ S_{xz}\ S_{yx}\ S_{yy}\ S_{yz}\ S_{zx}\ S_{zy}\ S_{zz}$.}
\item[restart]{This controls whether a checkpoint/restart file is read
in. Accepted values are (0) new simulation, do not read restart; (-1)
read a restart file but calculate fresh initial velocities; (1) use the
entire restart file to continue from a previous calculation}

\end{description}

\subsubsection{The system file}
The system file (by default \textit{system.in}) contains information on
the physical system including system size, shape and contents. An example
system file looks like this:

\begin{verbatim}
4000                 # Number of particles included in this file
1 1 1                # Number of replications of the system in x,y,z
56.2719571 0.0 0.0   # Size of the system volume (3x3)
00.0  32.4886349 0.0 #
0.0 0.0   51.8247969 #
    0.0333333351    0.0000000000    0.000000000    40  91.22
    0.0833333351    0.0500000000    0.000000000    40  91.22
    0.0666666651    0.0000000000    0.050000000    40  91.22
    0.0166666651    0.0500000000    0.050000000    40  91.22
.
.
\end{verbatim}

The first line contains the number of particles in the system file
itself. The simulation cell will be populated with a number of
copies of these particles, depending on the number of replications
entered on line two. Lines 3-5 are the unreplicated system volume size
and shape in Angstroms. After that follows $N_{particles}$ lines each giving the
fractional coordinates, the atomic number and mass of the particle.

The system file can be considered as defining a unit cell. The replication
feature allows a simulation to be constructed from any number
of replications of this unit cell, and is particularly useful for
creating large volumes of regular lattice positions.  Thus an
equivalent to the above file specifying all positions would be:

\begin{verbatim}
4                 # Number of particles included in this file
10 10 10                # Number of replications of the system in x,y,z
5.62719571 0.0 0.0   # Size of the system volume (3x3)
00.0  3.24886349 0.0 #
0.0 0.0   5.18247969 #
    0.333333351    0.0000000000   0.000000000    40  91.22
    0.833333351    0.500000000    0.000000000    40  91.22
    0.666666651    0.0000000000   0.500000000    40  91.22
    0.166666651    0.5000000000   0.500000000    40  91.22
\end{verbatim}

\subsubsection{Checkpoint/restart}

MOLDY allows full checkpoint/restart governed by the \verb|restart|
and \verb|nchkpt| keys in the parameter file
(Section~\ref{parameterfile}). The file contains the entire simulation
parameter set, particle positions, velocities, accelerations and
jolts, the atomic number, species, and mass index arrays and bulk
thermodynamic properties of the system. Checkpoint/restart files are
created every \verb|nchkpt| timesteps (set in the parameter file) and
at the end of a simulation.
The output file system.out can be used in place of \textit{system.in} to restart the
simulation.  It is also in a format readable by BallViewer \cite{33}.

\subsubsection{Tests and examples}

MOLDY includes several examples that may be used to help gain familiarity with the procedure for setting up and running simulations.  These examples are stored in the \verb!trunk/examples/! directory and are documented on the MOLDY wiki website: \url{https://www.wiki.ed.ac.uk/display/ComputerSim/MOLDY}. Tab.~\ref{tab:examples} lists the set of examples at time of writing; these may be added to in the future.

\label{tab:examples}
\begin{table}[h]
\caption{Set of examples included with the initial release of MOLDY.}
\begin{tabular}{lrl}
Test & $N_{particles}$  & Notes\\
\hline
\verb|Iron heat| & 2000 &  Bcc iron at constant pressure \\
\verb|Melting| & 16000 &  Zr constant pressure MD  at high temperature \\
\verb|Dislocation| & 3600 &  Bcc iron with line defect \\
\end{tabular}
\end{table}


\section{Results}

All the results presented in the following sections use one of
the available potentials described in Section~\ref{Available potentials}. The tests have
been run on a machine with four quad-core Intel Xeon E7310 processors
running at 1.6 GHz.

\label{section:parallel_performance}

\subsection{Parallel benchmarks}

The benchmark test case considered is that of a cube of bcc iron as generated by the following system input file:
\begin{verbatim}
2               # Number of particles included in this file
n n n           # Number of replications of the system in x,y,z
2.870 0.0 0.0   # Size of the system volume (3x3)
0.0 2.870 0.0   #
0.0 0.0 2.870   #
0.0 0.0 0.0 26 55.847		# Iron atom at the origin
0.5 0.5 0.5 26 55.847		# Iron atom at the centre of the cell
\end{verbatim}
We vary $n$ to create cubes of a given size as required.  Unless otherwise stated, benchmarks were performed at a temperature of 300 K with no thermostat enabled at a constant pressure of 0.1 GPa with periodic boundaries in all directions.  A timestep of 1 fs was used throughout and the system evolved for 1000 timesteps using the Verlet integrator.  Due to the lack of diffusion in our system we do not re-evaluate the neighbour list for the duration of the simulation.

The parallel performance of MOLDY was assessed on a shared memory machine with four quad-core Intel Xeon E7310 processors running at 1.6 GHz with 16 GB of main memory.  All calculations were carried out at full 64-bit precision.  Each benchmark is repeated three times to ensure that timing results are reliable.

\begin{figure*}
  \begin{center}
\subfigure[A plot showing how the runtime scales with the number of atoms in the system.  As expected linear scaling is observed.]{\label{graph:natoms_scaling}\includegraphics[width=0.8\textwidth]{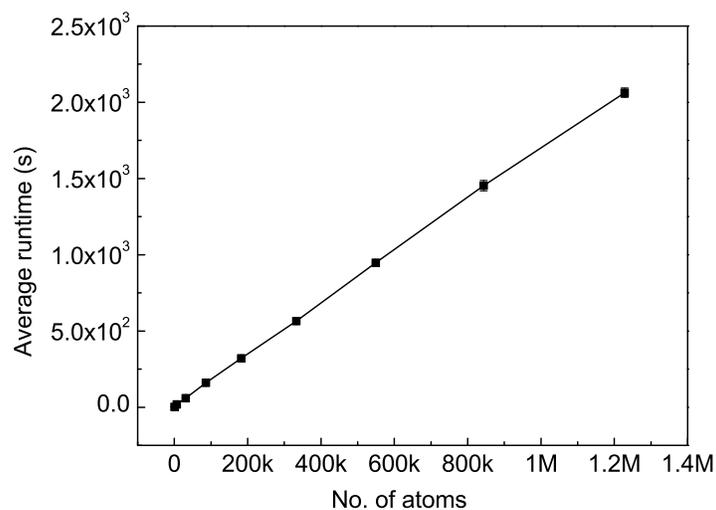}}
				\subfigure[A plot of the runtime per atom when using 16 threads.  This levels off at $\approx$2000 atoms per thread, any less and parallelisation overheads begin to impact on code efficiency.]{\label{graph:runtime_per_atom}\includegraphics[width=0.8\textwidth]{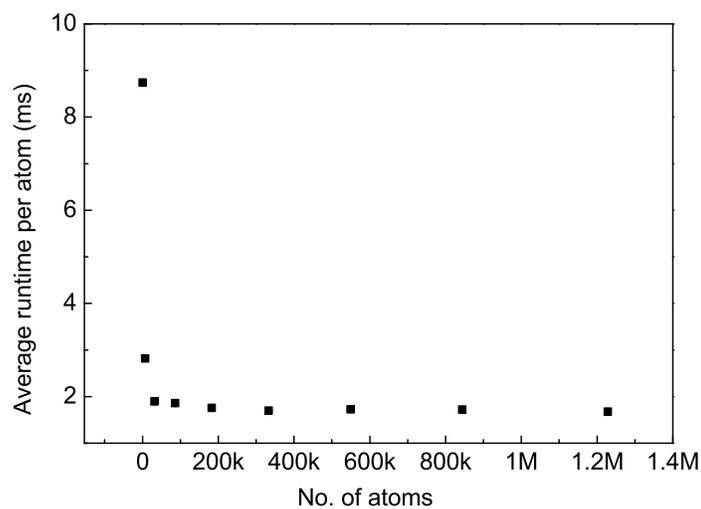}}
	\end{center}
  \caption{Graphs showing how MOLDY scales with the number of atoms in the system.}
  \label{graph:natoms}
\end{figure*}

Fig.~\ref{graph:natoms_scaling} shows how the runtime scales with the number of atoms in the system.  As expected, linear scaling is observed as a consequence of MOLDY's short-range potential nature.  This benchmark set was carried out using all 16 available threads and we can therefore also consider the runtime per atom which is plotted in Fig.~\ref{graph:runtime_per_atom}.  Initially starting at $\approx$0.013 s this levels off to $\approx$0.092 s when using 32k atoms or more.  This suggests that each processor requires at least 2000 atoms to become saturated with sufficient work during each iteration.

\begin{figure*}
  \begin{center}
\subfigure[Speedup comparison using two system sizes and lookup table versus direct potential evaluation. The iron potential used here is of intermediate complexity: a sum of 15 cubic splines.]{\label{graph:strong_scaling_speedup}\includegraphics[width=0.8\textwidth]{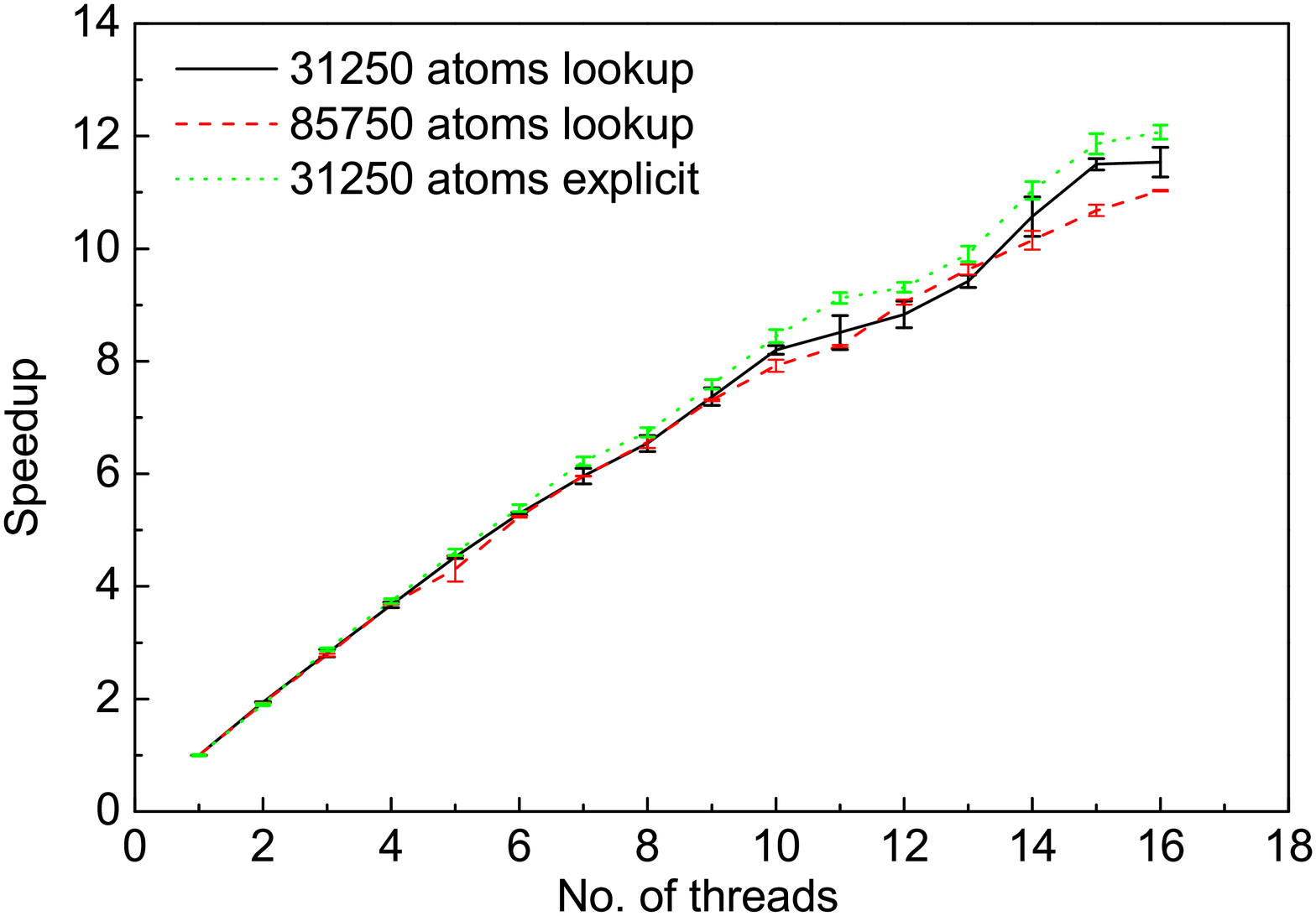}}
				\subfigure[Comparison in performance between using the lookup table and explicit potential evaluation.  Here we use a system of 31,250 atoms.]{\label{graph:strong_scaling_explicit}\includegraphics[width=0.8\textwidth]{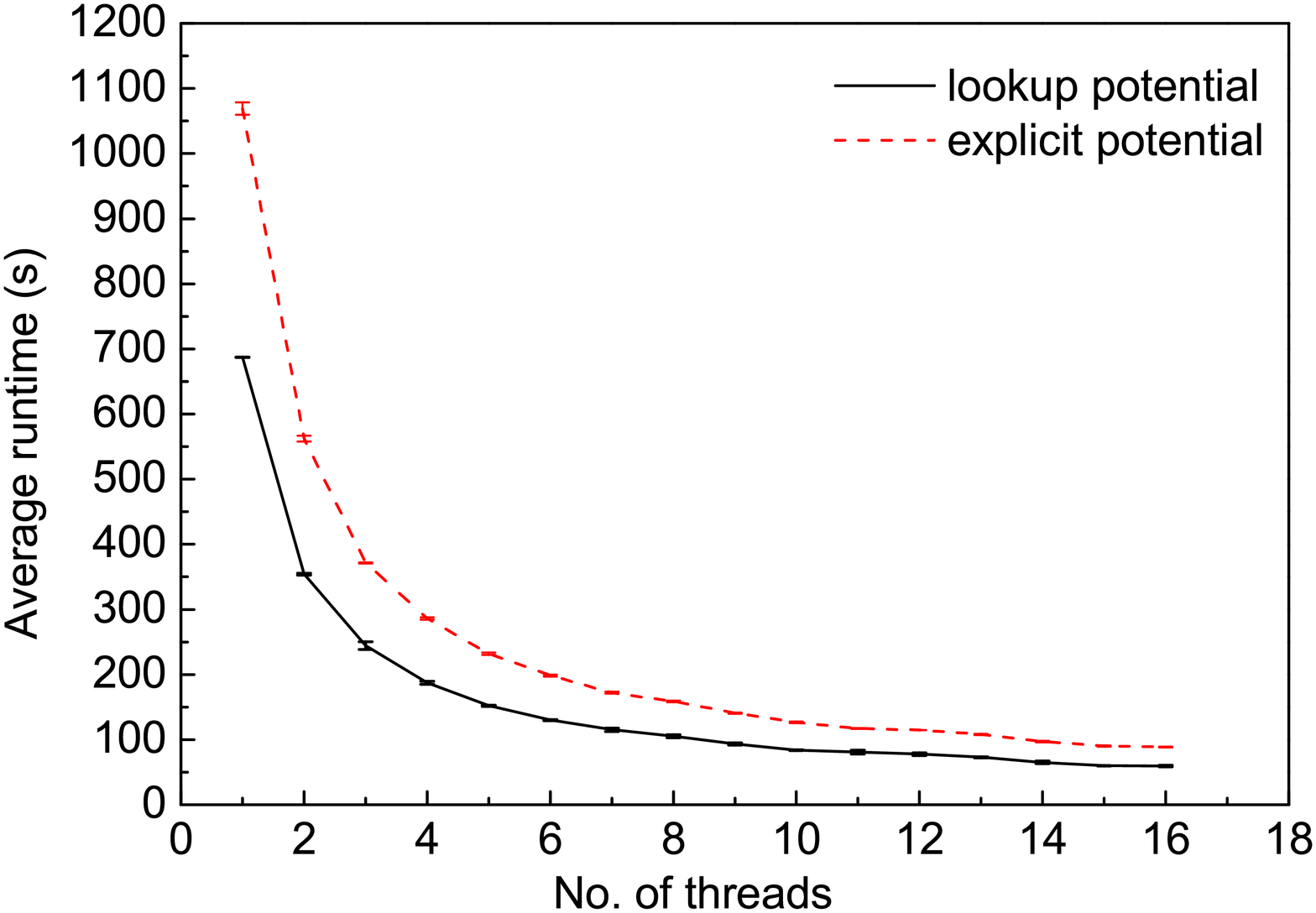}}
	\end{center}
  \caption{Parallel performance of MOLDY when integrating a system of bcc iron for 1000 timesteps.}
  \label{graph:strong_scaling}
\end{figure*}

We evaluate the strong scaling using systems of 32k ($n$=25) and 85k ($n$=35) atoms.  Fig.~\ref{graph:strong_scaling_speedup} shows the speedup when running using the lookup table and explicit potential calculation.  We see that good scaling is achieved all the way up to our maximum of 16 threads.  The explicit potential shows the best speedup; this is due to the fact that it takes longer to evaluate the force consequently increasing the parallel fraction of the code.  Fig.~\ref{graph:strong_scaling_explicit} confirms that, in absolute terms, the lookup table method significant outperforms the explicit potential evaluation even at 16 threads.

\begin{figure}
 \centering
		\includegraphics[width=0.8\textwidth]{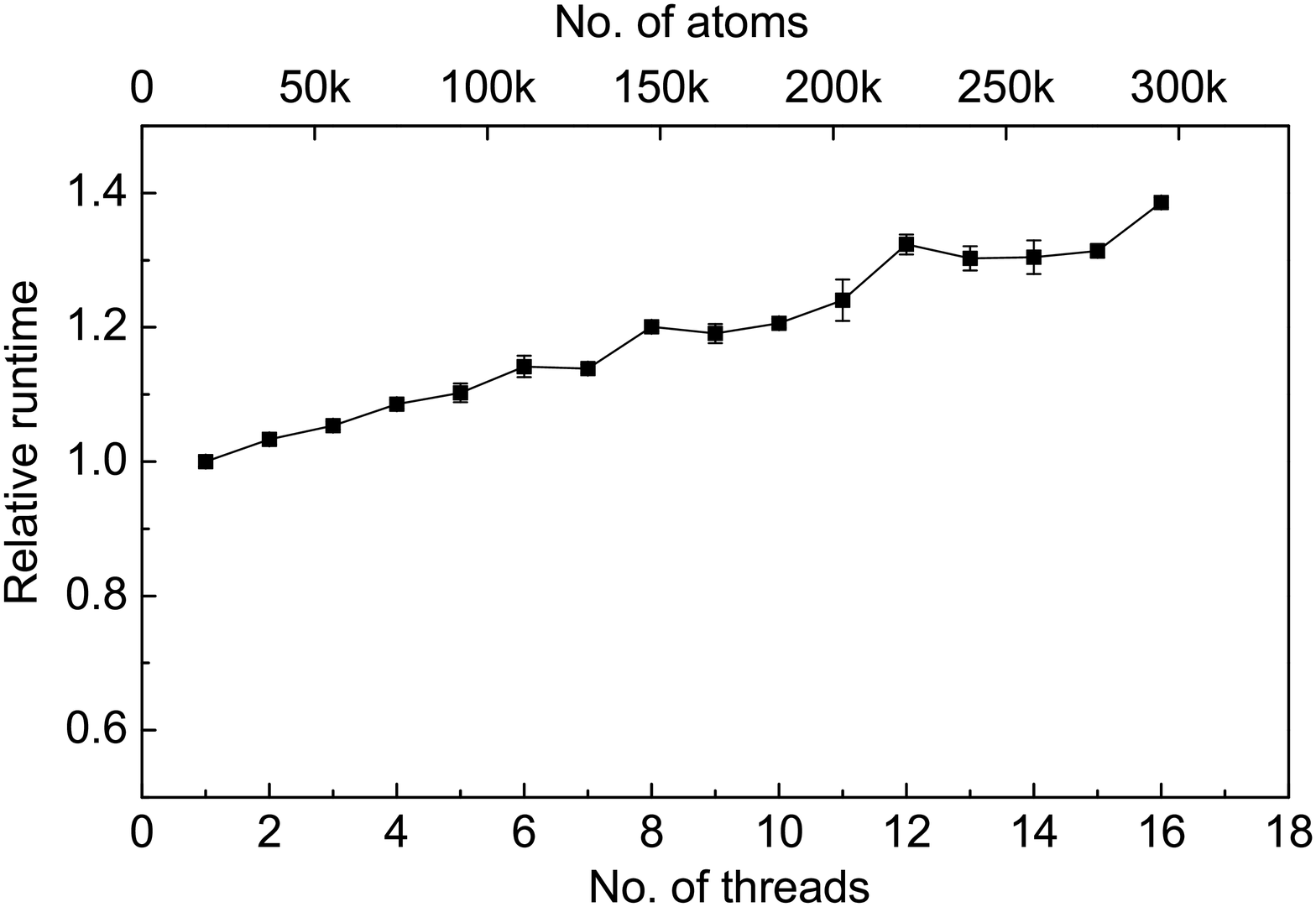}
	 \caption{The weak scaling performance of MOLDY plotted relative to the single thread runtime.}
  \label{graph:weak_scaling}
\end{figure}

To evaluate the weak scaling we start with a system of 18k atoms (24 $\times$ 24 $\times$ 16 repeats) and increase the system size linearly with the number of threads by scaling the z-repeats accordingly.  Fig.~\ref{graph:weak_scaling} shows a shallow rise in relative runtime for each additional thread caused by the serial fraction.  The achieved scaling is, however, very good with a system of almost 300k atoms taking only 1.4 times as long to simulate using 16 threads as that of a 18k atom system on one thread.

\begin{figure}
 \centering
		\includegraphics[width=0.8\textwidth]{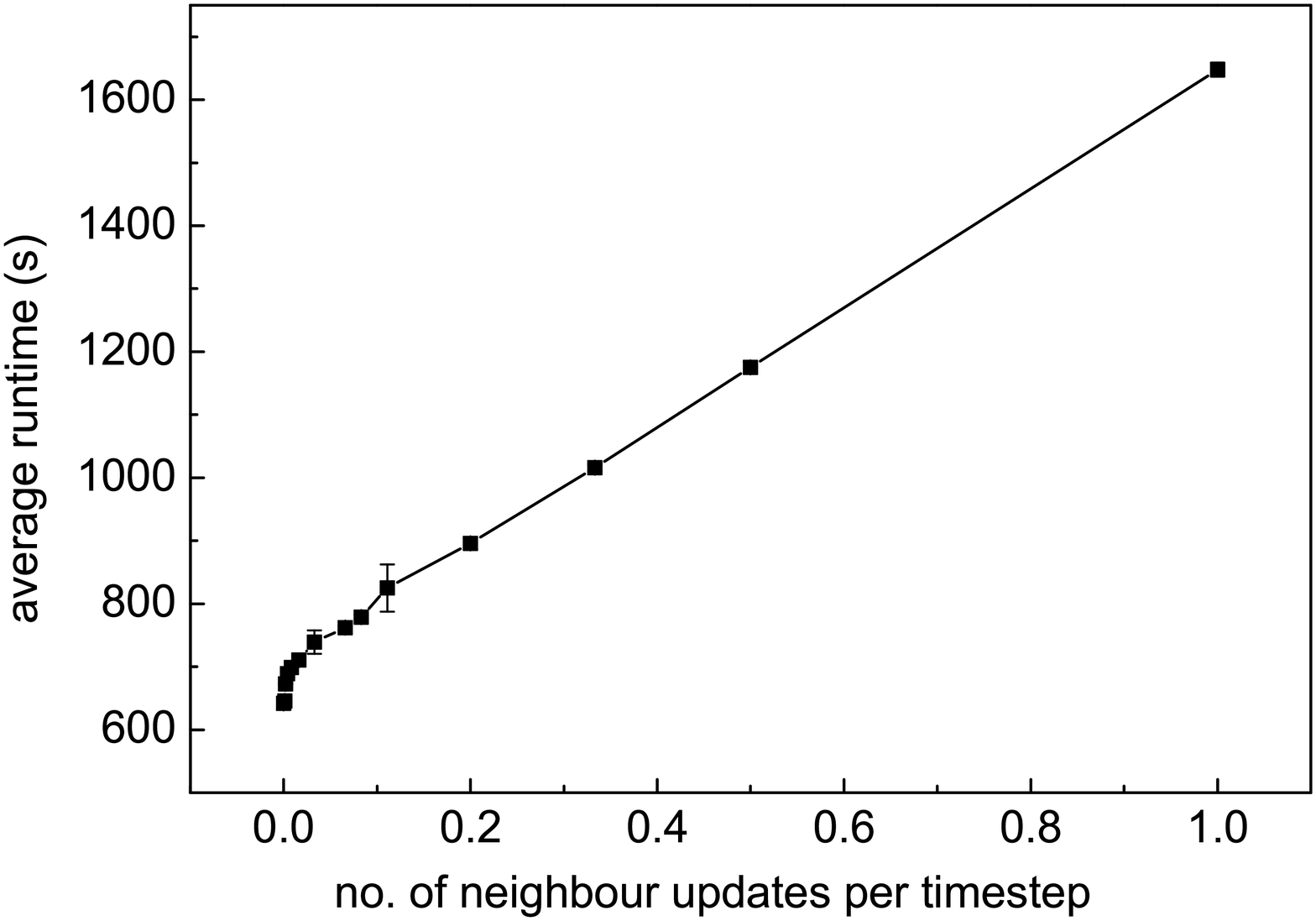}
	 \caption{The runtime scaling plotted as a function of the
	 neighbour list update frequency when running on one thread.
	 This implies that a full neighbour update takes almost twice as long as the force calculation}
  \label{graph:neighbour_updates}
\end{figure}

Finally, we evaluate the cost of neighbour list update frequency on the performance for a fixed number of threads.  This may become important for simulations with a high rate of diffusion where atoms can migrate between linked-cells often. Fig.~\ref{graph:neighbour_updates} shows how changing this parameter affects the runtime when using a single thread.  Initially the performance cost of increasing the update frequency is fairly steep, however this quickly relaxes to linear scaling with a fairly small scale factor. From this we recommend that for simulations with little or no diffusion neighbour list updates should be kept to a minimum while for simulations with large amount of diffusion it is safe to set a conservatively high frequency without incurring a large performance penalty.  It should be noted that the neighbour list update subroutine has not been parallelised and consequently a high neighbour list update frequency may adversely affect parallel speedup.

Further tests show that using a Nos\'{e} thermostat or predictor-corrector integrator in place of Verlet has little impact on the performance.  The Nos\'{e} subroutine is parallelised and has been found not to affect the parallel speedup.



\label{section:integration_error}
\subsubsection{Integration error}

\begin{figure}
 \centering
		\includegraphics[width=0.8\textwidth]{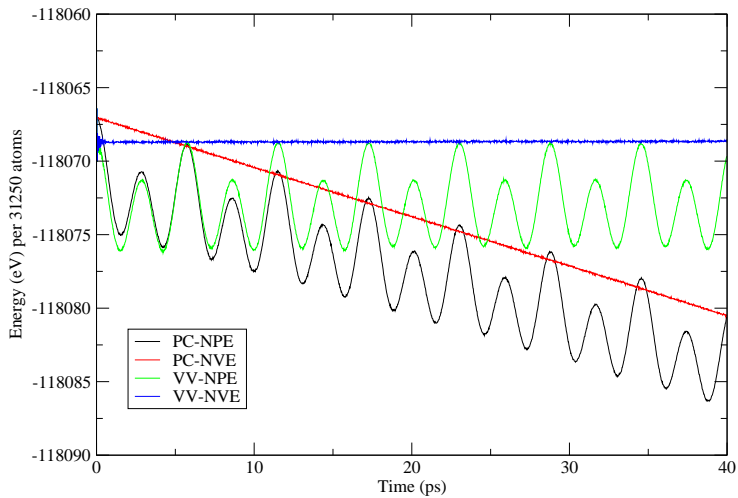}
	 \caption{The total energy plotted up to 20k timesteps of 2fs when using the Verlet and predictor-corrector integration schemes.  Both NVE and NPE ensembles are shown. 
The test system used was that from Section~\ref{section:parallel_performance} with 31250 atoms.}
  \label{graph:integration_error}
\end{figure}

To evaluate the energy drift caused by integration error we compare
results from both the Verlet and predictor-corrector
schemes. Fig.~\ref{graph:integration_error} shows the total energy
plotted over 20k steps when using the test system from
Section~\ref{section:parallel_performance} comprising 31250 atoms.  To
obtain some drift we used a 2fs timestep, slightly larger than one
would normally choose at such high temperatures.  We started each
simulation far from equilibrium, with atoms on their lattice sites and
1800K worth of kinetic energy.  Through equipartition the temperature
drops rapidly to around 900K.  The coupling to the box modes is weak,
and the NPE simulations oscillate in volume. In the figure, this
manifests itself as an oscillation in energy as energy is transferred
in and out of the fictitious modes.

Owing to its symplectic nature in the NVE ensemble, the Verlet scheme
has a bounded local error and preserves the total energy globally (see
e.g. \cite{10} for a discussion of integration error from Verlet).
For the NVT, NPT with Nose-Hoover thermostat no symplectic integrator
is possible and energy is not conserved.  For NPH ensembles Verlet is
not symplectic, and since energy is exchanged between the physical
system and the other terms in the Parrinello-Rahman extended
Lagrangian, energy conservation should not be expected.  In fact,
energy is very well conserved in the extended Hamiltonian: for the
physical system it oscillates about a fixed value.

Conversely the predictor-corrector exhibits a linear drift with
respect to the number of timesteps in NVE ensemble, with a similar
drift superimposed on the oscillations for the NPE case.  In this
instance this amounts to an error of $\approx$0.01\% over 20k steps.
While this error is small we recommend Verlet as the \textit{de facto}
integration scheme to be used especially in cases where thermodynamic
averages are the primary quantities of interest.


\section{Application Examples}

\subsection{Dynamic compression of copper}
One of the main features of MOLDY is its very flexible approach to
boundary conditions.  An example where boundary conditions are
important comes in simulating stamping a metal.  In this experiment,
the sample is subject to large uniaxial compression, inducing plastic
flow into free space in the perpendicular direction.

We simulate this process for a perfect nanocrystallite of copper, with
no preexisting dislocations, oriented along 100, cut along 001 and 010.
The boundary conditions applied are a
constant strain rate in the z-direction, with free boundaries in x and
y.  The process is fast, and the simulation region represents the
entire sample, so we do not use a thermostat.

The constant strain rate is applied using
$nsteps=1$ $strainloops=1000000$ and $STRX=0.9999995$ which applies a
Parrinello-Rahman box strain to the sample every timestep.
With a timestep of 1fs and a 108$\AA$~cell
this is equivalent to a realistic stamping velocity of $5.4m/s$.

Snapshots of the material at various levels of strain are analysed by
the BallViewer software \cite{33}. This shows a curious sequence of events.

The material transformation occurs first by elastic strain up to the
stress required for nucleation of dislocations.  This is significantly
higher than the yield stress of copper, implying that the barrier is
the nucleation rather than the motion of the dislocations.  It is
lower than the nucleation stress calculated for perfect single crystal
copper \cite{11} consistent with the apparent nucleation 
at the surface.

Partial dislocations then travel through the material leaving steps on
the surface: they leave behind them layers of material indexed as hcp
on account of its local stacking\cite{33}.  Because the dislocations
move on planes at an angle to the applied strain, they cross and lock,
hardening the material. Slip bands are formed and the stress
increases, until the macroscopic shape of the $yz$ plane changes from
a square to a diamond.  The orientation is now (111)
(11$\overline{1}$) (1$\overline{1}$0).  This change accompanies the
removal of most of the stacking faults associated with the
dislocations and indexed as hcp by BallViewer: the crystallite has
effectively recrystallised into its new shape.  This process is
associated with a rise in temperature due mainly to the work done in
compression.  The recrystallised sample has surfaces which are now the
(111) type, which have lower surface energy than the original (001)
type.  however the transformation does not create the lowest energy
Wulff shape: the steps on the surface from the previous dislocation
motion remain.

Further compression takes the system through another cycle of dislocation
generation, with the diamond shape becoming increasingly stretched.
This novel deformation mechanism would be strongly
suppressed by periodic boundary conditions.

The stress and potential energy of the sample is monitored and shown
in Fig.~\ref{fig:stress}.  The potential energy rises because energy is
being supplied to the system from work done in compression.  One can
estimate the T=0 potential energy by assuming equipartition of energy
between PE and KE, and evaluating PE-KE.  This quantity rises steeply
during the elastic phase, then drops, presumably due to reconstruction
at the surface, in the second phase.  In the third phase it increases
slowly, this is primarily due to the increased surface area and only
partly due to damage accumulating in the structure. Quenching the
structure to a local potential energy minimum followed by a BallViewer
analysis shows the final state to be almost perfect fcc.

\begin{figure}[ht]
  \begin{center}
   \includegraphics[width=5in]{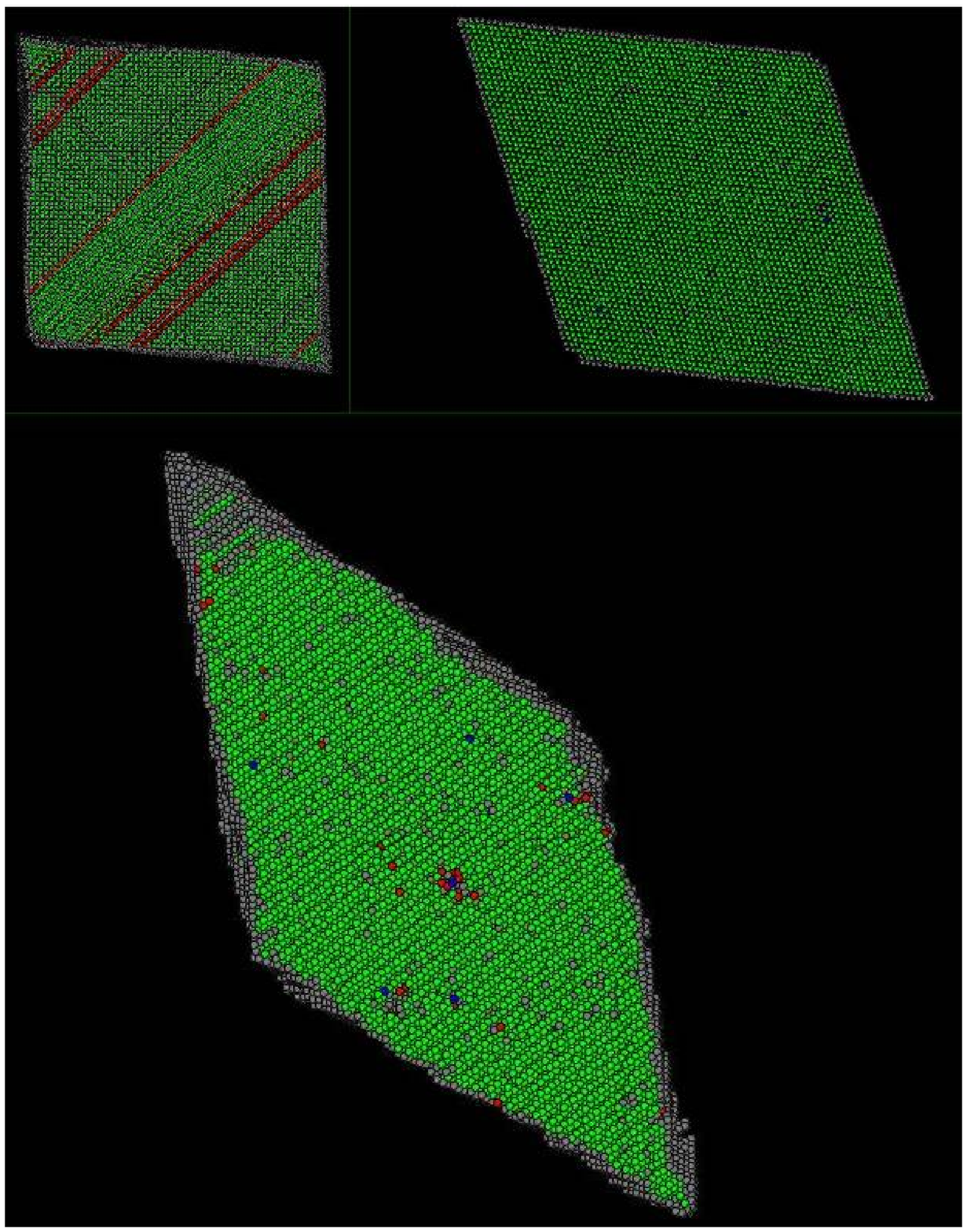}
  \end{center}\vspace{2mm}
\caption{Top view from BallViewer of stamping simulations, all at the same
scale.  Green atoms are fcc, red are hcp; top left, initial yield:
slip bands fringed with stacking faults; top right: reorientation
transformation complete, Bottom: final state after further dislocation
damage, note roughened edges.}\label{fig:BallView}
\end{figure}
\begin{figure}[ht]
  \begin{center}
   \includegraphics[width=5in]{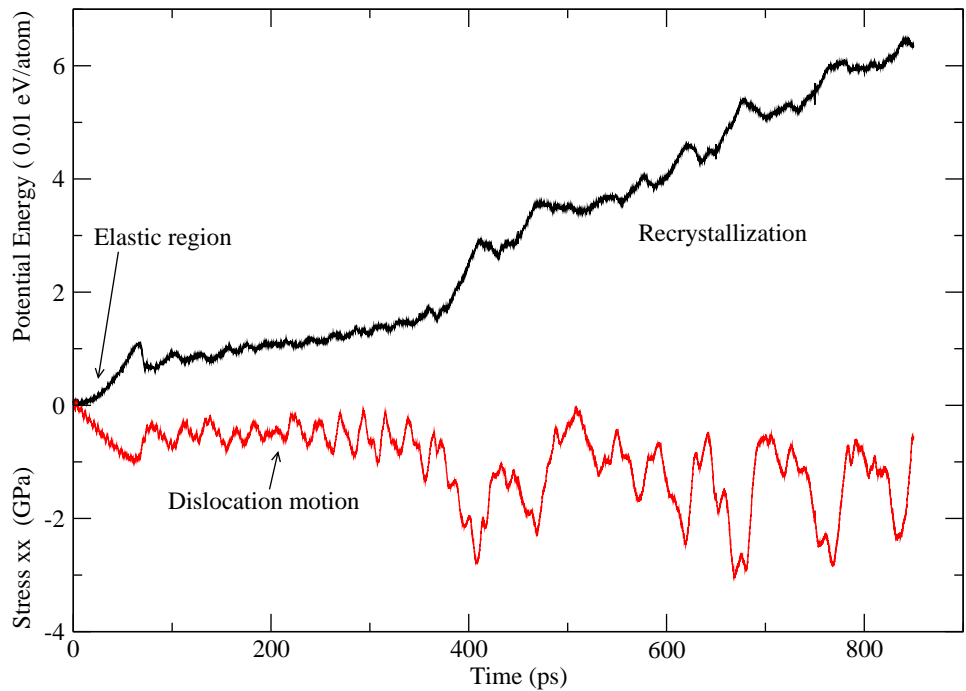}
  \end{center}\vspace{2mm}
\caption{
Evolution of stress and temperature in the constant energy constant strain rate
  ensemble stamping simulation. Regimes are shown corresponding to the
  initial elastic regime, the dislocation nucleation glide, and the
  reorientation/recrystallisation induced shape change of the
  sample.}\label{fig:stress}
\end{figure}
\subsection{Self-diffusion in iron}


Self-diffusion, D(T), can be related to the limiting slope of the mean square displacement (MSD) of the atoms
\begin{equation}
    D(T) = \lim_{t\to\infty}\frac{1}{6t} \langle \Delta{\bf r}(t) ^2\rangle
\end{equation}
where the term in the angle brackets is the MSD averaged over all
atoms and $t$ is the simulation time.  Self-diffusion for
$\alpha$-iron was measured for the temperature range from 900K to
2000K using a system of 2000 atoms with a single mono-vacancy at a
random site. The potential does not describe the ferro-paramagnetic
transformation, so the bcc structure is stable at all these
temperatures. The N-particle average of MSD against time
is shown in Fig.~\ref{diff}. Clearly, system diffusion increases at
higher temperatures, as expected. To characterise the temperature
dependency of diffusion and calculate the energy of a mono-vacancy
migration the data is fitted to the Arrhenius-type expression:
$D(T)=D_0 \exp(\frac{E_f+E_m}{k_B T})$, where $k_B$ is Boltzmann
constant, $T$ is temperature and E$_f$ and E$_m$ are the energies of
defect formation and migration, respectively.

The energy of vacancy migration was calculated to be E$_m^v$ = 0.60eV
with 99\% confidence limits of (0.53-0.67)eV. The fit excluded
temperatures $T > 1800K$, where the diffusivity increases
dramatically, hinting at a solid-liquid phase transition. This is in
reasonable agreement with the experimental value of the melting point
of iron; T$_{m}$~=~1811K. In addition, the energy of formation of a
vacancy was calculated using molecular statics: $E_f$ = 1.72eV. Both
of the these values are in good agreement with previous experimental
and simulation results presented in Tab.~\ref{diff}. The
prefactor $D_0$ was obtained from the y-intercept in
the Arrhenius graph to be $\nobreak{4.65\cdot10^{-3}cm^{2}/s}$ with
corresponding 99\% confidence bounds of
(2.58-8.48$\cdot10^{-3})cm{^2}/s$. This is two orders of magnitude
lower than the experimentally measured values, however in reasonable
agreement with the more recent MD simulations \cite{12,13}. 
Note that due to the extrapolation to infinite temperature and the exponential sensitivity
to $Q$, there are typically high errors associated with the
experimentally determined values.

There is a significant stochastic error in our calculations, as D(T)
was sampled at 100K intervals only, but such a large discrepancy
points to an inadequacy of the potential: this may be related to the
fact that the potential does not describe the transformation from bcc to fcc and back.

\begin{table}
\caption{Simulation and experimental results for vacancy properties of
$\alpha$ iron.}
\label{tab:diff}
\begin{tabular}{lccc}
Reference & $E_m^v$ (eV) & $E_f^v$ (eV) &$Q=E_m^v+E_f^v$ (eV) \\
\hline
&Simulation \\
Calder and Bacon (1993) \cite{15}  &0.91 & 1.83 &2.74\\ 
Domain and Becquart (2002) \cite{17}  &0.65 & 1.9 & 2.55\\
Fu et al. (2004) \cite{18}  &0.67 & 2.07 & 2.74 \\ 
Mendelev and Mishin (2009) \cite{12}  &0.60 & 2.18 & 2.78 \\
\emph{this work} & 0.60 &  1.72 & 2.32 \\
\hline
&Experiment\\
Schaefer et al. (1977) \cite{20}  & 1.28 & 1.60 & 2.86 \\
Vehanen et al. (1982) \cite{21}  & 0.55 & - & - \\ 
De Schepper et al. (1983) \cite{22}  &  - & 2.0& -\\ 
Romanov et al. (2006) \cite{23}  &  0.73 & - & -\\ 
\end{tabular}
\end{table}

\subsection{Elastic constants in iron}

We tested the shear boundary conditions by calculating the elastic
constants of iron.  Elastic constants vary with temperature in a way
which is often well described by the semi empirical expression

\[C(T) = C(0) - \frac{s}{exp(T_E/T)-1} \label{Varshni} \]

where $T_E$ and $s$ are empirical parameters\cite{24}. 
A resonant
ultrasound spectroscopy study \cite{25} 
suggested that the $C'=\frac{1}{2}(C_{11}-C_{12})$ elastic
modulus dropped by over 15\% between 0 and 500K. Equation \ref{Varshni}
then implied that it would go to zero above 2000K.  We calculated the
modulus from distorting a unit cell of 2000 atoms according to the C'
strain and evaluating the elastic constant from fitting the total
energy at each temperature to a quadratic in the strain.  This is
cross-checked with the stress-strain relation. Since the potential is
fitted to the zero temperature elastic moduli, good agreement at low
temperature is expected.  In Fig.~\ref{fig:Cprime} we show that
although the calculated C' softens, it does not do so as
much as in the experiment.  This suggests that the potential does not
fully account for the physics behind the softening.  The C' elastic
constant involves the same strain as the Bain transformation from bcc
to fcc, which is also not correctly reproduced.

\begin{figure}[ht]
  \begin{center}
    \includegraphics[width=1\textwidth]{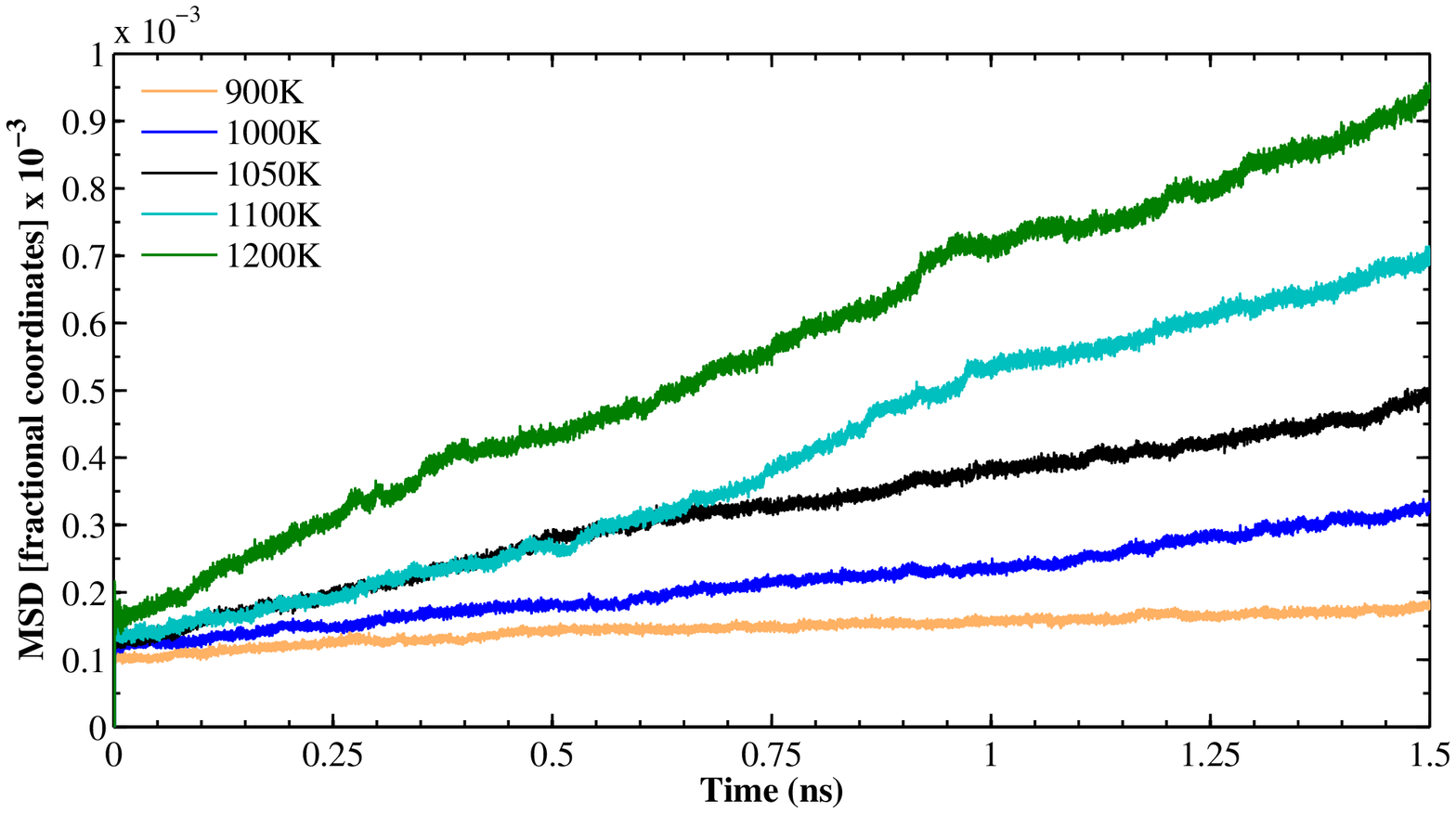}
  \end{center}
\caption{Calculated RMS displacements as a function of time and temperature
  from 2000 atom simulation with iron. Using the minimum image convention for
forces, but not wraparound for atomic positions, ensures that there are no
discontinuities when an atom leaves the supercell.}\label{diff}
\end{figure}

\begin{figure}[ht]
  \begin{center}
    \includegraphics[width=1\textwidth]{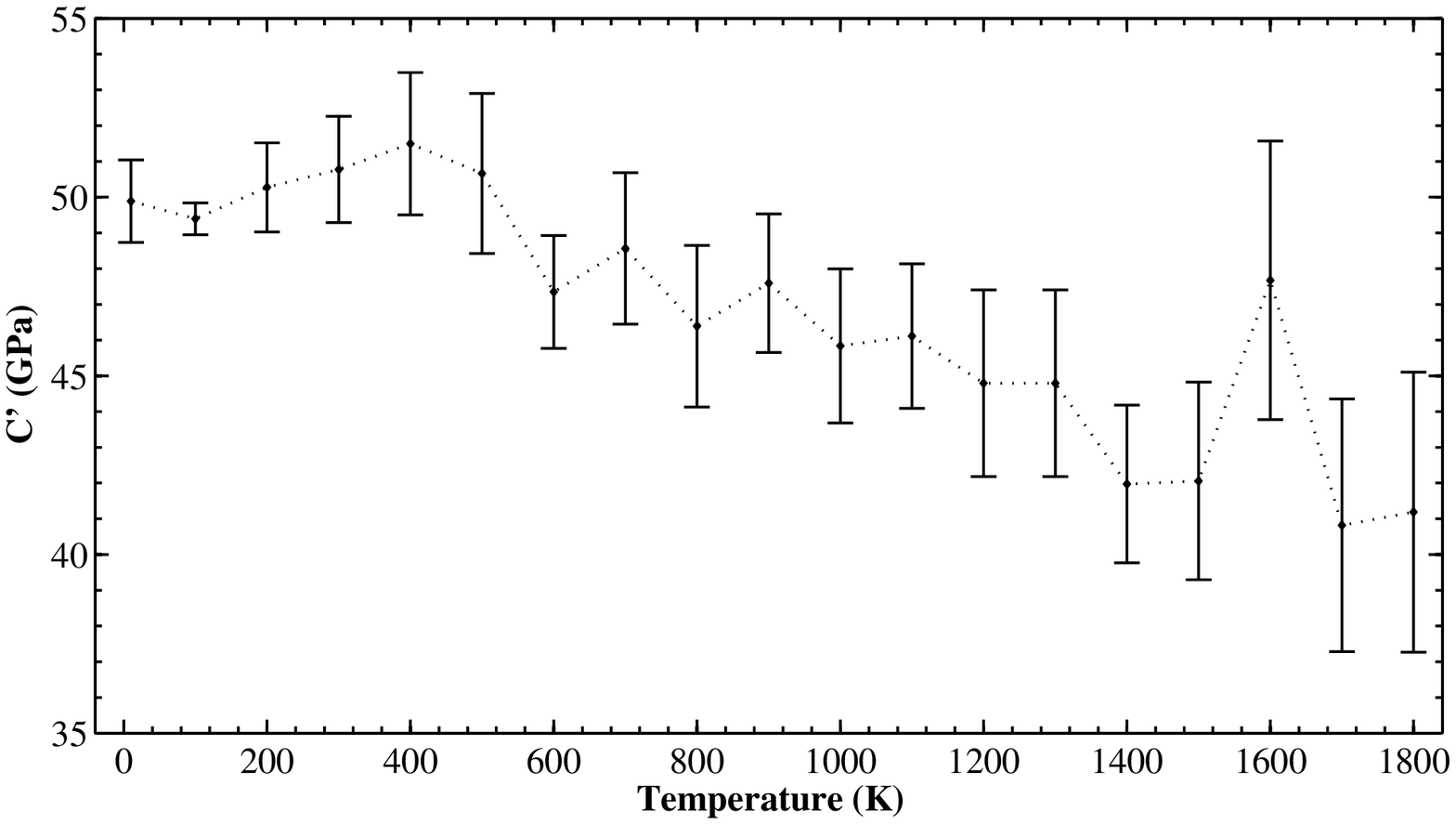}
  \end{center}
\caption{Variation of C' with temperature evaluated by applying finite
strains to the bcc structure and fitting a quadratic through the associated
energy.}\label{fig:Cprime}
\end{figure}

\subsection{Carbon migration in iron}

We also investigated the effect of carbon on diffusion in iron.

The diffusion D(T) temperature dependence and energy of migration for
carbon in bcc iron was evaluated for temperature ranging up to 1800K
with a 2000+1C atom cell.  The energy of migration of a carbon
interstitial was found to be considerably higher than that for the vacancy;
$E_{m}^{C} = 1.2(3)$eV$ > E_{m}^{v} = 0.6(1)$eV.
The carbon occupies the octahedral
$(1/2, 0, 0)$ site and the nearest adjacent site (e.g. $(1/2, 1/2,
0)$) lies along the (010)-type direction.  However,
Fig.~\ref{fig:C-diff} shows evidence of correlated migration along
various directions (e.g. event at 2500fs) and multiple correlated
motion along a (100) direction (e.g. event at 7000fs).  This suggests
that the martensite-type tetragonal strain field around the C can
persist, enabling secondary jumps to occur \cite{26}. 
\begin{figure}[here]
  \begin{center}
    \includegraphics[width=1\textwidth]{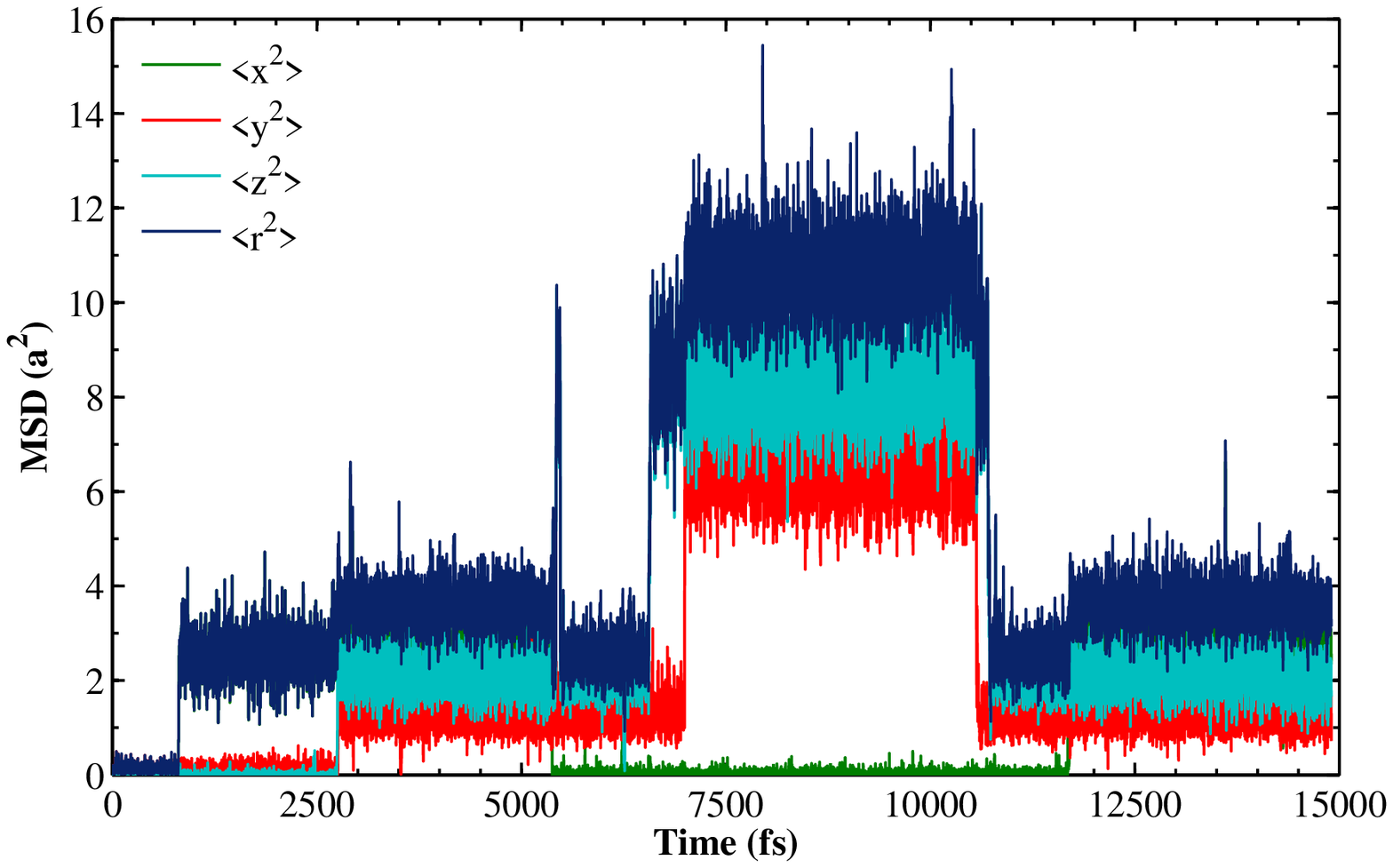}
  \end{center}
\caption{RMS displacement of the carbon atom through a series of jumps at 1300K.}\label{fig:C-diff}
\end{figure}

We also investigated a 1999+1C iron system with a carbon and a vacancy.  Now we
find a trapping and detrapping effect: the fastest migrating species
is the vacancy, followed by the vacancy-carbon complex and finally the
isolated carbon (see Fig.~\ref{fig:CV-diff}).  Thus the carbon acts as a weak trap for the vacancy,
slowing its migration: by contrast the vacancy enhances the
migration of the carbon.  Even the weak vacancy trapping effect means that
the overall self-diffusion rate of Fe is reduced by more than an
order of magnitude due to the presence of 0.05\% carbon.


\begin{figure}[ht]
  \centering
  \begin{minipage}[l]{0.49\textwidth}
    \centering
    \includegraphics[width=1\textwidth]{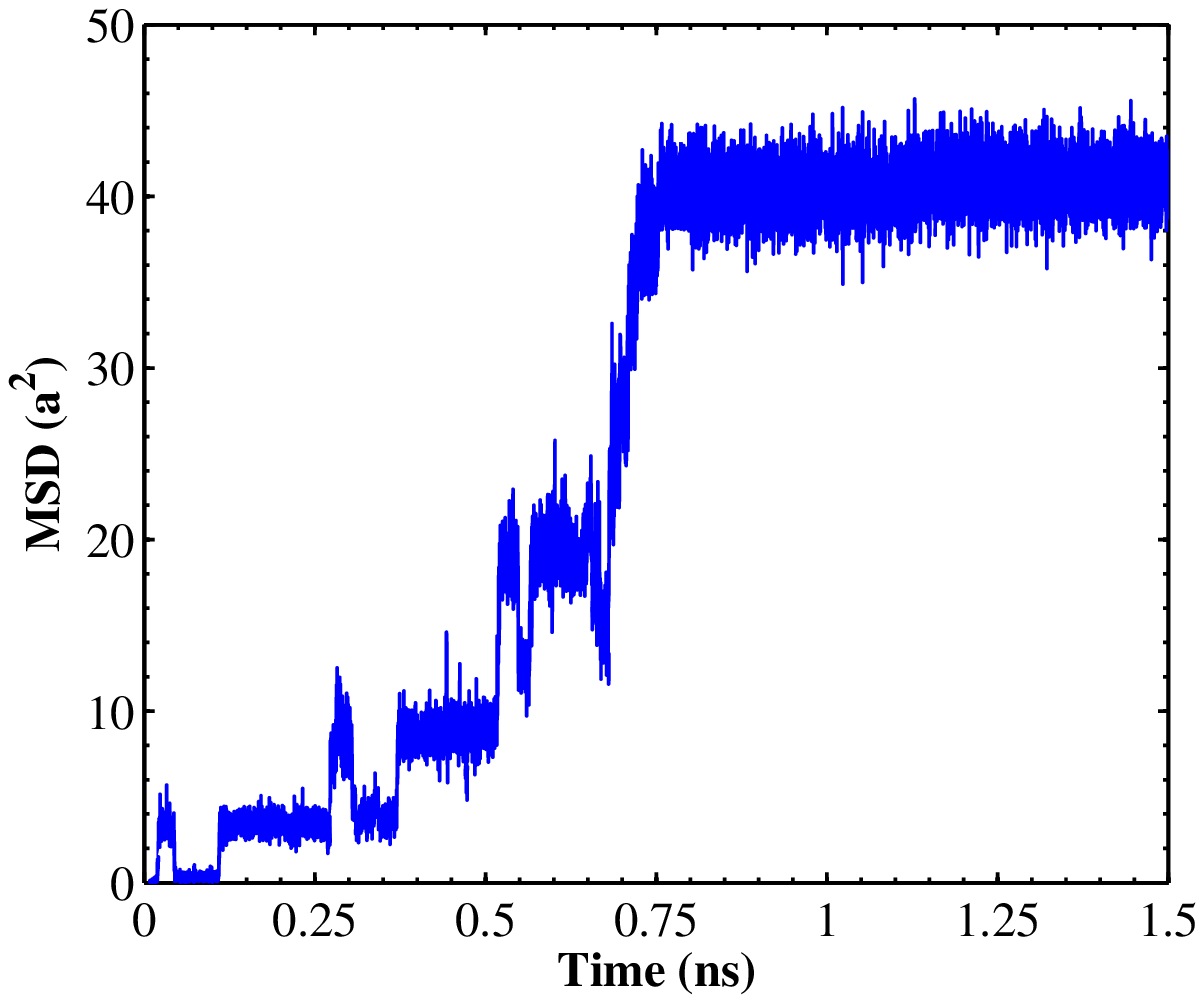}
  \end{minipage}
  \begin{minipage}[r]{0.49\textwidth}
    \centering
    \includegraphics[width=1\textwidth]{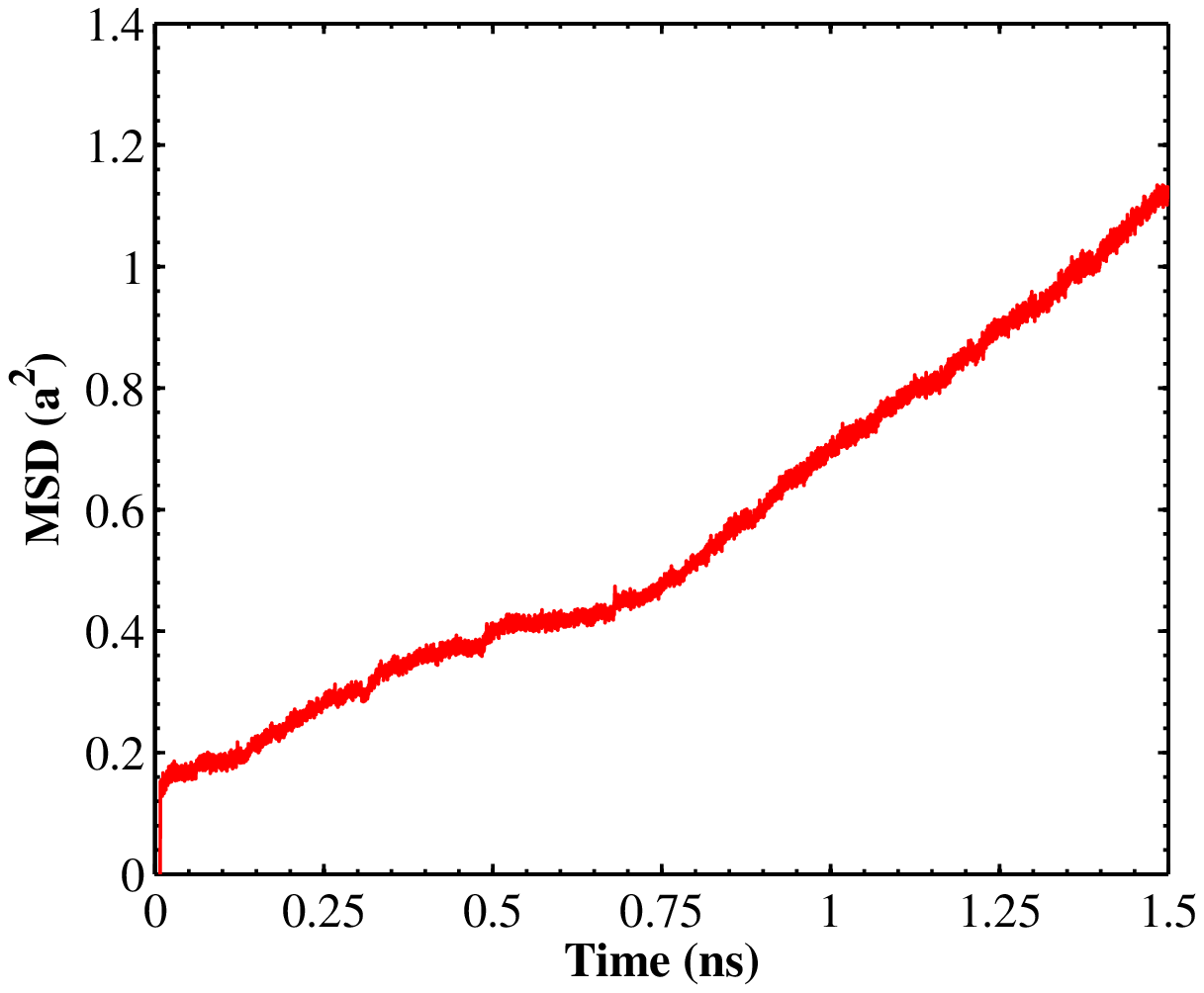}
  \end{minipage}
\caption{RMS displacement of the carbon atom through a series of jumps
in a system with a vacancy at 1350K.  Up to 800ps the carbon is
associated with the vacancy and migrates rapidly (left panel).  After
800ps carbon migration ceases, presumably due to
detrapping.  However migration of the vacancy is increased, as evidenced by
the increased iron self-diffusion (change in slope, right
panel).}
\label{fig:CV-diff}
\end{figure}

\section{Conclusions}

We described a thoroughly updated and parallelised flexible
molecular dynamics code MOLDY, which previously existed in many
different versions.  The program is interfaced with BallViewer, a bespoke
graphics package for identifying local crystal structures and
visualising MD outputs.

Furthermore, we demonstrated the parallel capabilities of MOLDY using up to 16 threads to run simulations of systems exceeding one million atoms.
It was shown that the strong and weak scaling of the code provided by OpenMP parallelisation were good out to a 16-thread system.

MOLDY comes with its own test suite and we have
presented applications, which illustrate some of the code features that are missing in many standard codes, such as variable boundary
conditions, quenching, autocorrelation, complete stress and strain
tensors. These applications show some features which are interesting
in their own right, for example the crystallographic reorientation of
a single nanocrystal of copper under uniaxial compression and the
100-fold reduction in iron self-diffusion due to a small amount of
carbon.

\section{Acknowledgements}

This work was supported by the EPSRC under grant number EP/F010680/1 and
by the EU-FP7 project GETMAT. We are grateful to Mark Bull for discussions
on OpenMP. We also thank our codehost http://code.google.com/p/moldy/ .






\end{small}
\end{document}